# Evidence of planetary influence on solar activity: Phase coherence of the variation in sunspot area with the tidal effect of Mercury.

I. R. Edmonds
12 Lentara St, Kenmore, Brisbane, Australia 4069.
Ph/Fax 61 7 3378 6586, ian@solartran.com.au

**Keywords :** planetary effect on sunspots; magnetic Rossby wave modes; periodic sunspot emergence

**Abstract.**
Reports of periodicities in solar activity in the intermediate period range between 11 years and 27 days, known as Rieger-type quasi-periodicities, have been numerous in the past 30 years. However, no generally accepted explanation for the episodic occurrence of the extensive range of Reiger-type periodicities has emerged. In this paper we examine the possibility that the periodicities are associated with a Mercury – Sun interaction of base period 88 days. To test this idea we band pass filter the 140 year long daily sunspot area data to obtain the 88 day period and 176 day sub harmonic period components of the data and compare the time variation of the components with the time variation of the orbital radius, $R_M$, of Mercury – or more specifically with the time variation of the factor $(1/R_M)^3$, proportional to the tidal effect of Mercury. We were able to show that, when successive episodes of the occurrence of the 88 day period component were discrete, i.e. not overlapping in time, the time variation of this component of sunspot area was either exactly in-phase or exactly in anti-phase with the time variation of $(1/R_M)^3$. A similar result was obtained for the 176 day period component. When several discrete episodes of the 88 day or the 176 day components occurred during a solar cycle the Fourier transform of the sunspot area data exhibited strong sidebands with periods dependent on the duration of the episodes. A simple model based on episode modulation and solar cycle modulation of 88 day and sub harmonic period sinusoids was able to reproduce most of the spectral peaks observed in the intermediate range of sunspot area periodicity. This is compelling evidence of a link between the motion of Mercury and the periodic emergence of sunspots. It is proposed that the link involves the existence of magnetic Rossby waves with mode periods close to the sub harmonic periods associated with Mercury and the triggering of sunspot emergence by those Rossby waves.

## 1. Introduction

A model accounting for the ~11 year cycle of sunspot emergence was described by Babcock (1961) as follows. The Sun, at solar minimum, has a polar magnetic field and no sunspots. Faster solar surface rotation near the equator wraps the subsurface lines of magnetic field in a longitudinal spiral around the Sun. This amplifies the magnetic pressure in the ropes of sub surface magnetic flux circling the Sun. Eventually increasing magnetic pressure causes the gravitational internal pressure of flux ropes to become lower than the gravitational external pressure. The flux ropes become buoyantly unstable and loops of flux rope rise through the solar surface in each hemisphere to form bipolar magnetic regions, each with two sunspots of opposite polarity, marking the onset of a new ~11 year solar cycle. As the cycle progresses sunspots in each hemisphere with polarity opposite to the polar field in that hemisphere migrate to the poles and begin the process of reversing the polarity there. The other sunspots in each hemisphere migrate to the equator and neutralise. Eventually the polar magnetic field is restored with opposite polarity and the equatorial region is again free of sunspots. This completes a Schwabe cycle, ~11 years, and one half of a Hale cycle, ~22 years. It is not clear why the Schwabe cycle is ~11 years long. Historically, and recently, connection to the motions of Jupiter and Saturn has been suggested, (Charbonneau 2002, Charbonneau 2013).

Periodicities in solar activity, are also observed in the intermediate period range between the ~11 year period Schwabe cycle and the ~ 27 day period solar rotation cycle. These are known as quasi-periodicities, Reiger (1984) due to their intermittency and variable periods. The quasi-periodicities occur in episodes of 1 – 3 years duration around the maximum of each solar cycle with periodicity around 150 days often observed, (Lean 1990, Richardson and Cane 2005, Ballester et al 2004). However the range of intermediate periodicity reported extends from 40 days to several years, Chowdhury et al 2015, Tan and Chen 2013). These enigmatic periodicities have been observed in nearly all solar activity and interplanetary space variables. In particular in flares, (Reiger et al 1984, Dennis 1985, Ichimoto et al 1985, Landscheidt 1986, Bai & Sturrock 1987, Droge et al 1990, Kile & Cliver 1991, Bai 1994, Bai 2003, Dimitropolou et al 2008, Bogart & Bai 1985); in proton events, (Gabriel et al 1990); in photospheric magnetic flux, (Ballester et al 2002, Knaack et al 2005, Rouillard and Lockwood 2004); in radio bursts Verma (1991); in sunspot number and area, (Lean & Brueckner 1989, Lean 1990, Pap et al 1990, Carbonell & Ballester 1990, Carbonell & Ballester 1992, Verma and Joshi 1987, Verma et al 1992, Oliver and Ballester 1995, Oliver et al 1998, Ballester et al 1999, Krivova & Solanki 2002, and Chowdhury et al 2015); in interplanetary magnetic field, Richardson & Cane (2005); and in cosmic ray flux (Mavromichalaki et al 2003, Dragic et al 2008, Hill et al 2001, Wang and Sheeley 2003 and Rouillard and Lockwood 2004). While there are a wide range of periodicities reported the most common reference is to the "~150 day quasi-periodicity", Richardson & Cane (2005). Zaqarashvili et al (2010) reported time/period diagrams showing quasi-periodicities in sunspot area for cycles 19 to 23. The diagrams show a broad range of periods and indicate strong episodes of periodicity in the period range 150 to 190 days occurring around the maximum of solar cycles 19 and 21. There is, currently, no accepted explanation for intermediate range quasi-periodicities.

Proposed explanations for quasi-periodicities include Bai et al (1987b) suggesting that the cause must be a mechanism that induces active regions to emit flares, Ichimoto et al (1985) suggesting that the quasi-periodicities are associated with the storage and escape of magnetic flux from the Sun, Bai & Cliver (1990) suggesting the periodicities could be simulated with a non-linear damped oscillator, (Bai & Sturrock 1991, Sturrock & Bai 1993) suggesting that the Sun contains a "clock" with a period of 25.5 days and the periodicities are sub-harmonics of the "clock" period, Wolff (1983, 1992, 1998) suggesting generation via the interaction of the solar activity band with solar g modes, Sturrock (1996) suggesting the Sun contains two "rotators", one at ~22 days and the other at ~25 days, that combine to produce the observed periodicities, Lockwood (2001) suggesting variations of ~ 1 year duration may be related to oscillations close to the base of the solar convection zone, Wang and Sheeley (2003) demonstrating that periodicities in the intermediate range in solar magnetic flux can occur through the random development of sunspot groups on the solar surface, and (Lou 2000, Lou et al 2003, Zaqarashvili et al 2010) suggesting the periodicities in the intermediate range can be linked to equatorially trapped magnetic Rossby waves near the surface of the Sun.

Rossby waves can form in a fluid layer on the surface of a sphere. Lou (2000) and earlier Wolff (1998) derived dispersion relations for equatorially trapped Rossby waves in the photosphere of the Sun. The amplitude of equatorially trapped Rossby waves vary longitudinally around the equator and have a Gaussian envelope spanning $60^{o}$ across the equator. Lou (2000) derived the following expression for the allowed periods:

$$T(p,q) = 25.1[q/2 + 0.17(2p + 1)/q] \text{ days} \quad (1)$$

where 25.1 days is the sidereal period of surface rotation of the Sun at the equator, p = 1, 2 and q = p, p+1, p+2 … The p = 1 mode has one nodal line through the pole, the p = 2 mode has two nodal lines through the poles. Figure 1 shows the allowed mode periods for the p = 1 and p = 2 modes as a function of q.

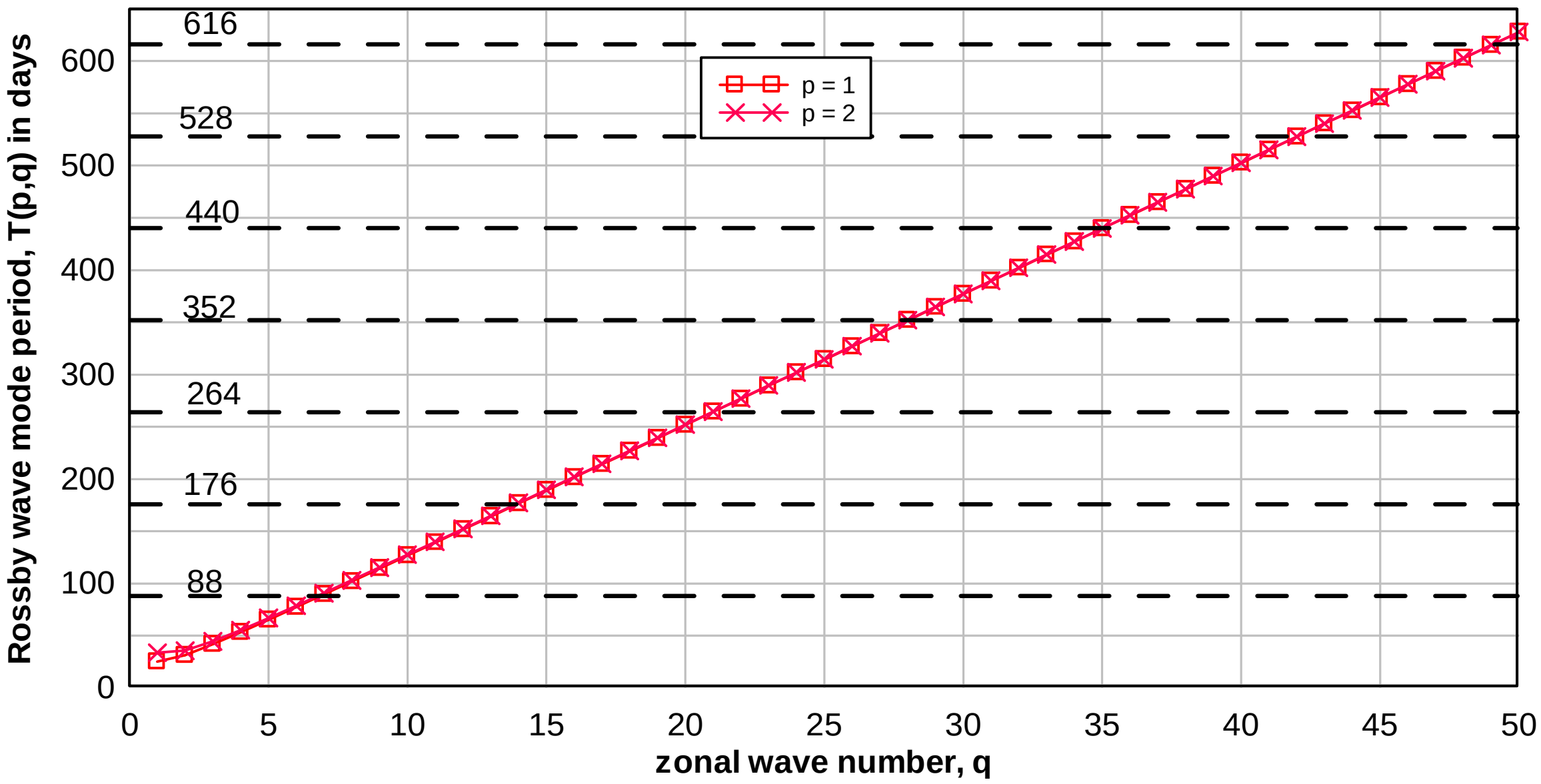

**Figure 1.** The magnetic Rossby wave period from equation (1) for nodal line numbers p =1 (squares) and p = 2 (crosses) as a function of wave number q. Horizontal lines indicate the period and sub harmonic periods of Mercury.

For modes at q = 5 and above there is a very small difference between the periods of the 1,q and the 2,q modes and mode periods can be found by T(p,q) = 25.1q/2. The reference lines in Figure 1 indicate that there is almost exact correspondence between the sub-harmonic periods of Mercury and every seventh Rossby mode period. The exact correspondence occurs because the period of Mercury, 88 days, is 3.50 times the period of photosphere rotation at the equator, 25.1 days. As a result the rotation of the equatorial photosphere is locked to the orbital motion of Mercury in the sense that after two orbits Mercury, at closest approach, is above the same point on the solar surface. A Rossby wave moves around the surface of the Sun at an angular speed, relative to the surface, of $2/q^2$ times the angular speed of the suns surface, Lou (2003). So for higher q number modes a Rossby wave is, essentially, a stationary wave on the solar surface. As Mercury, at closest approach when the tidal effect is strongest, is above the same point every second orbit it follows that any influence of Mercury is particularly suited to the long term stimulation of stationary Rossby waves with mode periods close to periods associated with Mercury. Dimitropoulou et al (2008) noted that the possible Rossby wave modes are quite dense, with spacing between allowed periods of about 12 days, so that many observed Reiger type quasi-periodicities could reasonably be attributed to Rossby waves on the Sun. In their concluding remark, Dimitropoulou et al (2008), noted that "an additional mechanism should be considered, which practically promotes specific modes out of the theoretically equivalent ones."

In this paper we consider the possibility that the "additional mechanism that promotes specific modes" is the tidal effect of the planet Mercury. The tidal effect is taken to be the height difference between high and low tide induced on the surface of the Sun by the gravitational effect of a planet. The height difference or tidal elongation due to a planet of mass M and orbital radius R at the surface of the Sun is $(3/2)(r_{Sun}^4/M_{Sun})(M/R^3)$ where $r_{Sun}$ is the radius of the Sun and $M_{Sun}$ is the mass of the Sun, Scafetta (2012). Thus any planetary tidal influence can be expected to vary as $1/R^3$. The orbit of Mercury is elliptical and, as the orbital radius of Mercury, $R_M$, varies by 40% from 0.47 to 0.31 astronomical units (A.U.) during an orbit, $1/R_M^3$ varies by 120% during each orbit of 88 days. This is the largest percentage variation of tidal effect among the planets. In the following we associate the tidal effect of Mercury with the variation of the quantity $(1/R_M)^3$ which varies from 9.8 A.U.$^{-3}$ to 34.4 A.U.$^{-3}$ during an orbit with average level 19.6 A.U.$^{-3}$.

The concept that periodic activity on the Sun may be associated with the periodicity of the planets is controversial as the calculated tidal height variations on the Sun are minute. The tidal height difference due to Mercury varies from 0.60 mm to 0.17 mm during an orbit and is therefore insignificant compared with ambient fluctuations at the solar surface or solar tacholine, (De Jager and Versteegh 2005, Callebaut et al 2012, Scafetta 2012). The history of the planetary tidal effect concept has been reviewed by Charbonneau (2002) and a revival of the planetary hypothesis has been discussed by Charbonneau (2013). The predominant idea by proponents is that tides on the Sun due to the planets somehow stimulate or trigger solar activity and the period of this resultant solar activity is similar to the period of a planet or to some more complex periodicity due to the conjunction or lining up of two or more planets, (Hung 2007, Scafetta 2012, Abreu et al 2012, Tan and Cheng 2013). Scafetta (2012) has proposed that the minute tidal effect could be greatly amplified by a mechanism involving increased conversion of mass into energy according to Einstein's equation when the tidal effect causes minute density changes in the interior of the Sun. The amplitude of the tidal effect due to Jupiter is similar in amplitude to the tidal effect due to Mercury. The period of Jupiter is 11.9 years and Scafetta (2012) has shown that, in combination with Earth and Venus, the combined tidal effect due to the three planets is phase coherent with the observed ~11 year solar cycle. The concept of a planetary effect on solar activity remains controversial due to the lack of a convincing mechanism that would allow minute tidal surface height variations to influence magnetic activity on the Sun with research continuing e.g. Wolff and Patrone (2010), Charbonneau (2013).

There are, nevertheless, several reasons for considering a planetary effect due to Mercury.

(1) Bigg (1967), using techniques from radioastronomy for detecting periodic signals buried in noise, showed that daily sunspot numbers for the years 1850 to 1960 contained a small but consistent periodicity at the period of Mercury which is partially modulated by the positions of Venus, Earth and Jupiter.

(2) As discussed earlier and illustrated in Figure 1, the coincidence of Rossby wave mode periods with the sub harmonic periods of Mercury. For example, the ratio between the period of the 1,28 mode, 351.85718 days, and the second sub harmonic period of Mercury, 351.87704 days, is 0.99994. Thus if the "additional mechanism promoting modes" is the tidal effect of Mercury one might expect to see evidence of Rossby wave modes on the Sun at ~ 88 days, and at the sub harmonic periods, ~ 176 days, ~246 days, ~ 352 days, etc.

(3). There is direct observational evidence of Rossby waves on the Sun with periods close to the periodicities associated with Mercury. Sturrock and Bertello (2010) provided a power spectrum analysis of 39,000 Mount Wilson solar diameter measurements between 1968 and 1997. They fitted Rossby wave mode frequencies to the eight strongest peaks in the spectrum. The two lowest frequency peaks in their tabulation were observed at frequencies 4.04 years$^{-1}$ and 2.01 years$^{-1}$. The corresponding day periods, at 90.4 days and 182 days, are each just 3% longer than, respectively, the 88 day period of Mercury and the 176 day period of the first sub harmonic of Mercury. Their observations therefore support the existence of Rossby wave modes on the Sun at periods very close to the periodicities associated with Mercury.

(4). Dominant periodicities in solar activity in the intermediate range are occasionally observed very close to the sub harmonic periods of Mercury. Lean (1990) found the strongest peak in periodograms of daily sunspot area occurred at 353 days. This period is 1.003 of the third sub harmonic of the period of Mercury, 351.88 days, and is also very close to the p =1 q = 28 Rossby wave mode period, 351.86 days, see equation 1. The periodicity of 1.68 years, (614 days), in magnetic flux examined by Rouillard and Lockwood (2004) is 0.996 of the period of the sixth sub harmonic of Mercury, 616 days. Tan and Chen (2012) found the strongest periodicity in the F28 microwave solar emission record was at 900 days. This is 1.02 of the ninth sub harmonic period at 880 days. Wang and Sheeley (2004) found the dominant peak in the spectra of solar magnetic field in solar cycle 21 was 1.7 years (621 days) and in solar cycle 22 was 2.9 years (1059 days). The periods are, respectively, 1.008 of the sixth sub-harmonic (616 days) and 1.003 of the eleventh sub-harmonic (1056 days).

(5). Like all planetary motion, the phase of the tidal effect of Mercury is predictable exactly so that it is possible to establish if the time variation of the examined periodicity in solar activity is in phase with the time variation of the planetary motion. Establishing phase coherence provides a higher level test of connection over and above that obtained by comparing periods obtained by spectral analysis. For example, Wang and Sheeley (2004) demonstrated that the spectral peak at 1.7 year periodicity in solar magnetic flux during cycle 21 might have arisen through random processes associated with sunspot evolution over a solar cycle. However, if it could be shown that the time variation of the 1.7 year component of solar magnetic flux during cycle 21 was phase coherent with the sixth sub harmonic of the time variation of the Mercury tidal effect this would provide added weight to the possibility of a connection.

To summarise: Mercury exerts a minute tidal force on the Sun with period ~88 days, Scafetta (2012). Theoretical estimates of Rossby wave periods, Lou (2000), predict allowed modes that, we note, have periods very close to the periodicities associated with Mercury. There is direct observational evidence, Sturrock and Bertello (2010), of Rossby waves on the Sun that, we note, have periods very close to the period and the first sub harmonic period of Mercury. Rossby waves may trigger the emergence of sunspots on the surface of the Sun, (Lou et al 2003, Zaqarashvili et al 2010). However, as a mechanism which promotes specific modes is required, Dimitropoulou et al (2008), we consider in this paper the possible influence of Mercury.

The arrangement of the paper is as follows: Section 2 outlines the method of Fourier analysis and band pass filtering. Section 3 describes the result of band pass filtering the sunspot area data between 1876 and 2015. Section 4 examines phase coherence between the tidal effect of Mercury and the ~ 88 day and ~ 176 day variations in the sunspot area. Section 5 outlines how the spectral content of the sunspot area data can be related to the periodicity of Mercury. Section 6 is a Discussion. Section 7 is the Conclusion.

## 2. Methods and data.

We use the Fast Fourier Transform (FFT) in the application D-Plot to detect periodicities in sunspot area data. FFT is the primary method of analysis used by other researchers to detect the Reiger type “quasi-periodicities” in the very large range of solar, space and terrestrial variables mentioned earlier. We also filter the sunspot area data with band pass filters centred on periods of 88 days and 176 days and compare the variation of the filtered data with the variation of the tidal effect of Mercury, specifically with the quantity $1/R_M^3$. The purpose of the comparison is to assess phase coherence between the variation of the filtered sunspot area data and $1/R_M^3$. The phase of $1/R_M^3$ is known precisely at any time. Therefore, if variation in sunspot area is related to the tidal effect of Mercury, it should be possible to demonstrate phase coherence of the band passed components of sunspot area and $1/R_M^3$ for intervals of tens or hundreds of years.

The method of band pass filtering to obtain the ~ 88 day period and ~176 day period components of sunspot area was as follows. A FFT of the entire sunspot area data series was made. The resulting n = 75 Fourier amplitude and phase pairs, $A_n(f_n)$, $\phi_n(f_n)$, in the frequency range 0.0102 days$^{-1}$ to 0.0125 days$^{-1}$ (period range 98 to 80 days) were then used to synthesize the ~ 88 day period band pass filtered version of sunspot area, denoted 88SSAN, by summing the x = 75 terms, $SSAN = A_x \mathrm{Cos}(2\pi f_x t - \phi_x)$ for each day in the series. Similarly, Fourier pairs in the frequency range between 0.00483 days$^{-1}$ and 0.00653 days$^{-1}$ (period range 207 days to 153 days) were used to synthesize the ~ 176 day period band pass filtered version of sunspot area denoted 176SSAN. Where data has been smoothed by, for example, a 365 day running average, the resulting smoothed data is denoted by the suffix Snnn e.g. a 365 day running average of sunspot area North data would be denoted SSAN S365.

In the present study there are two main variables: the orbital radius of Mercury, $R_M$, and the daily sunspot area on the northern hemisphere of the Sun (SSAN). SSAN is measured by the Greenwich Observatory in units of the area of one millionth of a solar hemisphere or microhems. The data is available at http://solarscience.msfc.nasa.gov/greenwch/daily_area.txt
The data begins in 1874. However, due to gaps, we use data from January 01 1876.
Daily values of the orbital radius, $R_M$, of Mercury in astronomical units (A.U.) are available at http://omniweb.gsfc.nasa.gov/coho/helios/planet.html for 1959 to 2019. Outside this range past values of the orbital radius of Mercury were calculated using

$$R_M = 0.38725 - 0.07975\cos[2\pi(t - 24.5)/87.96926] \quad \text{A.U.} \qquad (2)$$

where time in days, t, is measured from 0 at January 01, 1995.

**3. The ~ 88 day and ~ 176 day components of sunspot area North, 1876 to 2015.**
The amplitudes of the ~88 day period and ~ 176 day period filtered components of SSAN are shown in Figure 2. In what follows, for brevity, we will use e.g. the term "~88 day component" to refer to "~88 day period component". The ~176 day component in Figure 2 is displaced by -600 microhem for clarity. Also shown, the 365 day running average of the daily SSAN and numbers indicating the number of the solar cycle.

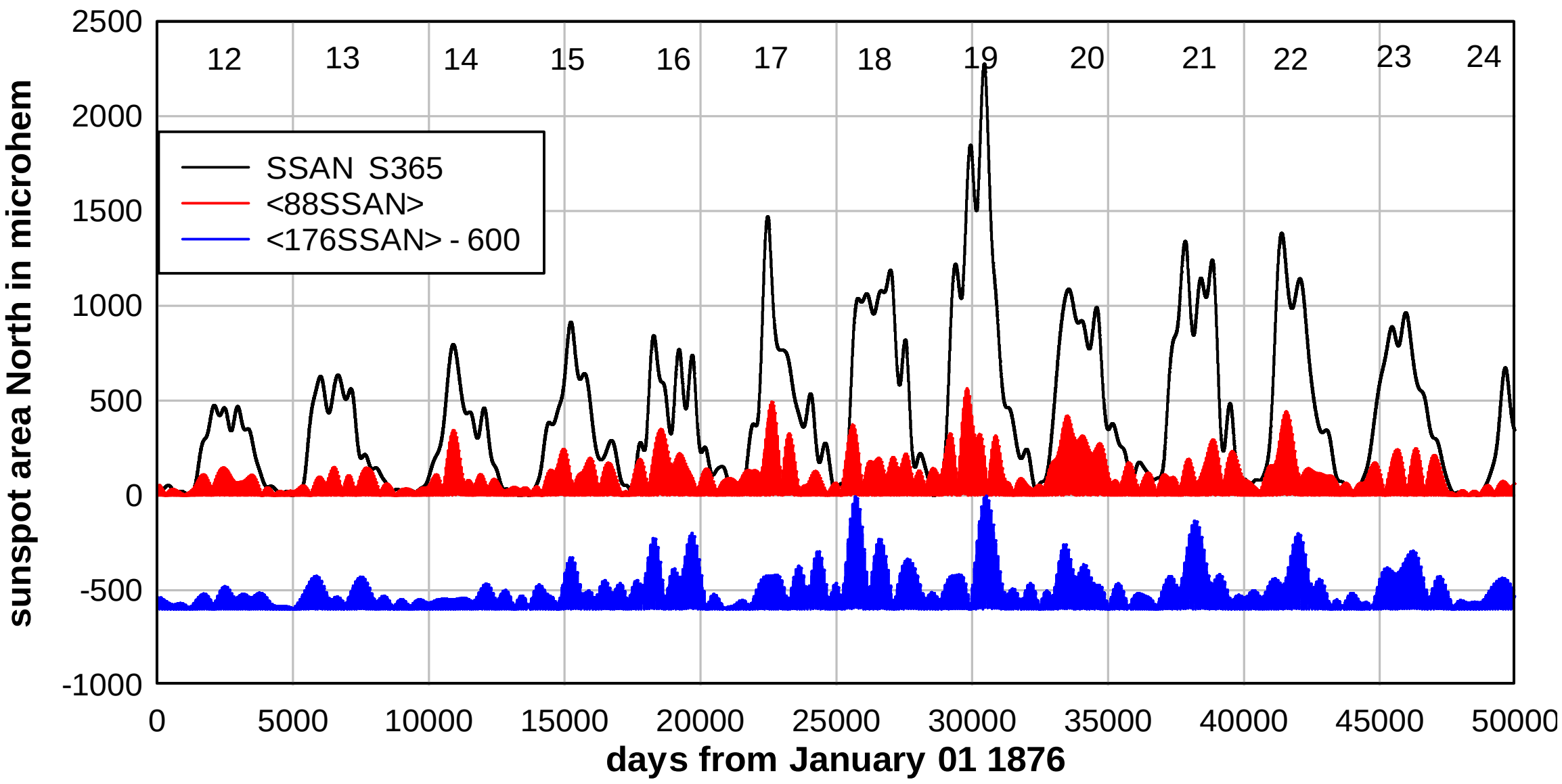

**Figure 2.** The amplitude of the ~ 88 day period (red) and the ~ 176 day period (blue) components of sunspot area North from 1864 to 2015. Also shown the 365 day running average of SSAN, (black).

It is clear from Figure 2 that variations of the ~88 day and ~176 day components are episodic. The shortest episodes last about 1.5 years e.g. each of the four episodes of the

~88 day component during cycle 23. The longest episodes last about 5 years e.g. the ~176 day component during cycle 19. Occasionally the episodes are discrete, i.e. not overlapping with adjacent episodes, as in cycle 23. However, more often episodes partly overlap e.g. the three episodes of the ~88 day component in cycle 20. We will show in the next section that, when an ~88 day episode is discrete, the variation of the ~88 day component of sunspot area during the episode is phase coherent with the variation of the tidal effect of Mercury or, more specifically, is exactly in-phase or exactly anti-phase with $1/R_M^3$. Similarly, when a ~176 day episode is discrete the ~176 day variation, during the episode, is exactly in-phase or anti-phase with the first sub harmonic of the tidal effect of Mercury.

We infer that the episodes shown in Figure 2 correspond to intervals of emergence of magnetic flux onto the surface of the Sun in the form of sunspots. We suggest that Rossby waves at 88 days period and 176 days period trigger buoyant instability of sub surface magnetic flux resulting in pockets of buoyant magnetic flux periodically floating up through the convection zone and emerging as sunspots in the photosphere. This process continues in a dynamic feedback cycle similar to that described by Lou et al (2003) until buoyant magnetic flux in that region of the Sun is exhausted and the episode ends. We suggest that sometimes only one region of the Sun is influenced by this process and the observed episode is discrete. At other times two similar period Rossby wave modes are involved, two or more regions of the Sun are simultaneously responding, and two or more episodes of sunspot emergence overlap in time.

It is clear from Figure 2 that the combined effect of the episodes in the ~88 day and ~176 day components is similar to the pattern of peaks at the top of each sunspot area cycle. For example, the twin peak pattern at the top of cycle 22 may be due to the strong ~88 day episode followed by the strong ~176 day episode. Similarly, the very square top form of solar cycle 18 may be due to the five near equal amplitude and short ~88 day episodes in combination with the three near equal amplitude and discrete ~176 day episodes in this solar cycle. We infer from the results in Figure 2 that the shape of each sunspot area cycle is largely the result of the combined effect of the separate ~88 day component and ~176 day component episodes of sunspot emergence during the cycle.

**4. Phase coherence of the ~88 day and ~176 day components with $1/R_M^3$.**

From the pattern of episodes in Figure 2 it is clear that ~88 day episodes of sunspot emergence usually overlap to some extent. Also the amplitude of the ~88 day episodes varies strongly within a solar cycle and varies strongly over the entire record. For example, the amplitude in solar cycles 12 and 13 is markedly lower than the amplitude in later solar cycles. In this section we are interested in examining if phase coherence exists between the variation of the ~88 day component of sunspot area and the 88 day variation of $1/R_M^3$. It is important to note that the average amplitude of the ~88 day component of SSAN in Figure 2 is 77 μhem which is a significant fraction of the average level of the SSAN data, 430 μhem. Thus if all episodes of the ~88 day component were in-phase with the 88 day tidal effect of Mercury a sharp and very significant peak at 88 day period would be evident in the spectrum of the SSAN record. However we find only minor peaks at 89.36, 88.56 and 87.94 days. Similarly, Lean (1990) found no significant peak at 88 day period in SSAN data for solar cycles 12 to 21. However, it is well known from the field of communication engineering that, when a signal of period T is phase modulated so that a $\pi$ phase shift or a time shift of T/2 occurs between regular intervals of the signal, spectral power is shifted into sidebands and there is a minimum of spectral power at period T. Thus, absence of spectral power at period T does not necessarily imply absence of causation at period T. We discuss this concept in detail in the next section where sunspot area data is examined by spectral analysis. However, we note here that the concept of phase modulation has previously been used to interpret the observed $\pi$ shift in Reiger type periodicity between solar cycles, Bai and Cliver (1990).

Figure 3 shows the ~ 88 day component of sunspot area North during 11 years of solar cycle 23. Also shown is the variation of $(1/R_M^3)$ from its average value, denoted $\Delta(1/R_M^3)$. Solar cycle 23 was chosen for examination as Figure 2 shows the ~88 day episodes of sunspot emergence in this solar cycle are strong and discrete.

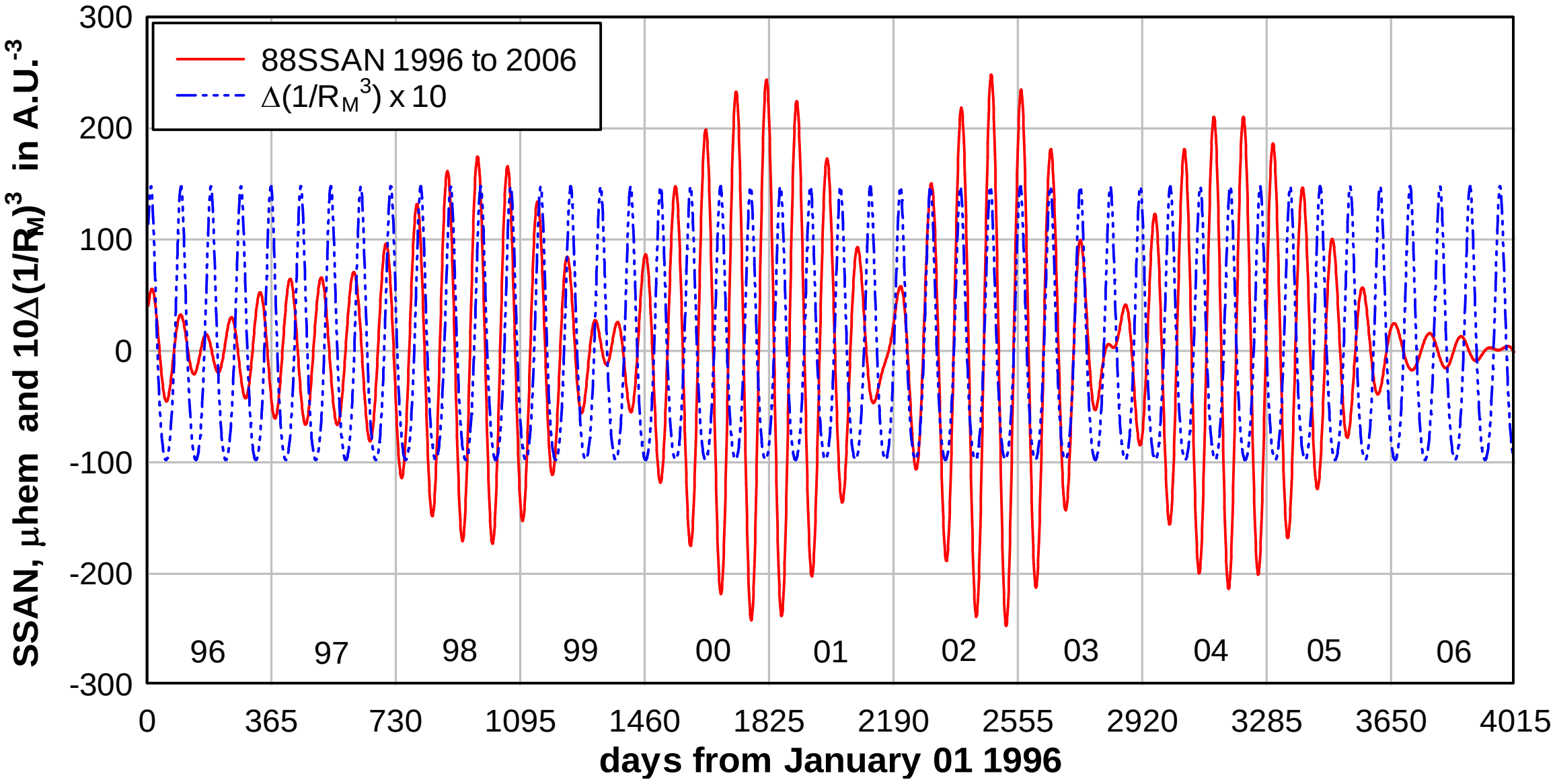

**Figure 3.** The ~ 88 day component of sunspot area North for 1996 to 2006 in solar cycle 23 (full line) compared with the variation of the tidal effect of Mercury, $\Delta(1/R_M^3)$. One weak and four strong episodes of the ~88 day component of SSAN are evident with the four strong discrete episodes showing exact in-phase or exact anti-phase coherence with $\Delta(1/R_M^3)$.

There are five episodes of the ~ 88 day component of SSAN during cycle 23. A weak episode in 1997 overlaps the first strong episode in 1998. The strong episode in 1998 is followed by three strong and discrete episodes; the first in 2000 2001, the second in 2002 2003 and the third in 2004 2005. During the four strong episodes it is evident that the phase coherence between the ~88 day variation of sunspot area North and $\Delta(1/R_M^3)$ alternates between exact in-phase coherence and exact anti-phase coherence. We infer from this result that when the episodes are discrete succeeding episodes occur with a $\pi$ phase shift relative to the preceding episode. Why this $\pi$ phase shift between sequential discrete episodes occurs is not clear. However, as the allowed periods of 1,7 and 2,7 Rossby wave modes are both very close to 88 days, see Figure 1, it may be that successive episodes of sunspot emergence respond alternately to p = 1 or p = 2 Rossby wave modes and this may account for the observed π phase shift. However, we note that, allowing for the π phase shifts between the sequential discrete episodes, the ~88 day component of sunspot area North is exactly phase coherent with the tidal effect of Mercury from 1998 to 2005, an interval that encompasses 33 orbits of Mercury.

Figure 4 shows the ~176 day component of sunspot area North and $\Delta(1/R_M^3)$ during 11 years of solar cycle 23.

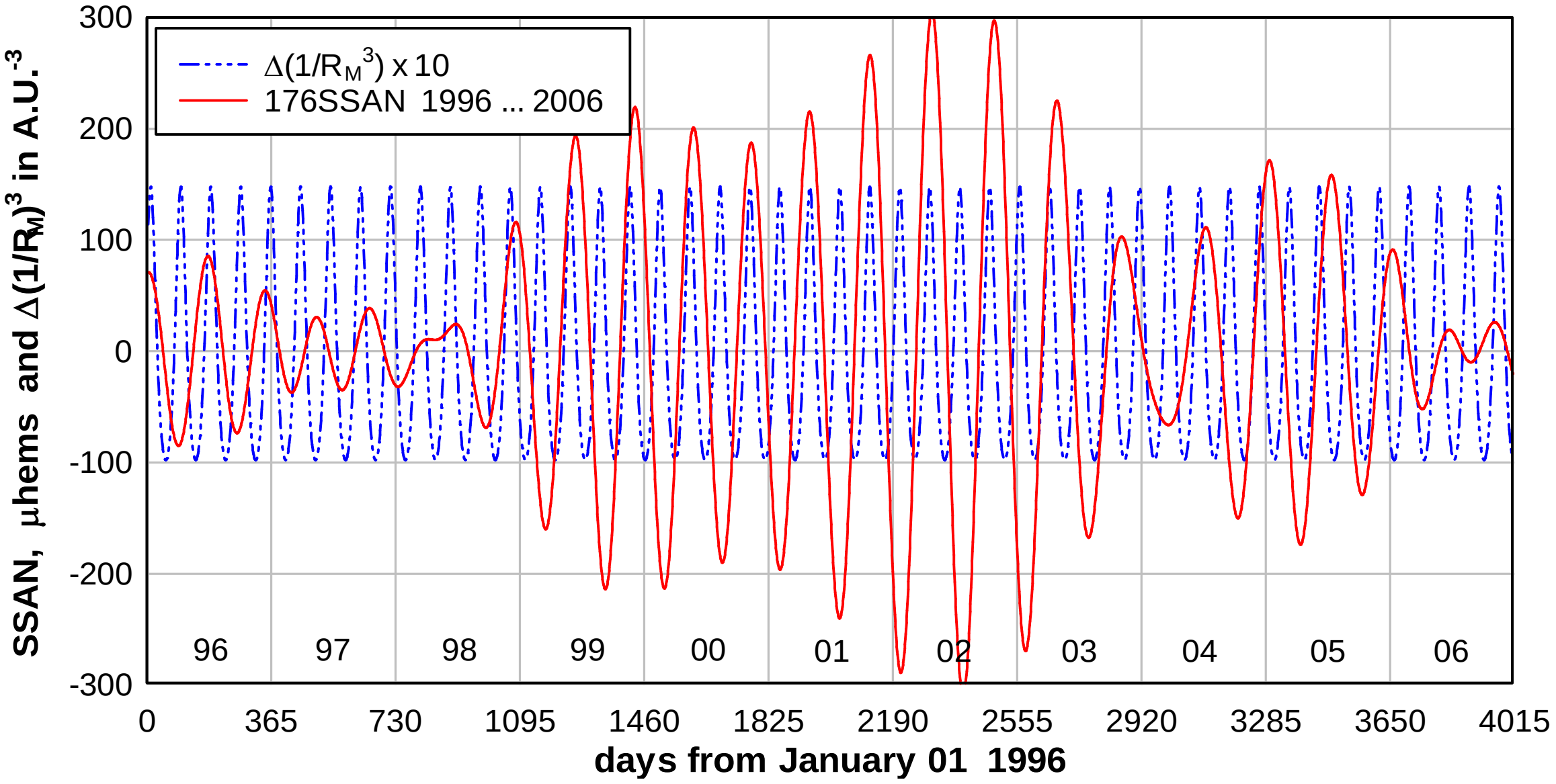

**Figure 4.** The ~176 day component of SSAN for 1996 to 2006 in solar cycle 23 (full line) compared with the variation of $\Delta(1/R_M^3)$. There is one dominant episode of ~176 day sunspot emergence during this interval in phase with every second peak of $\Delta(1/R_M^3)$.

There is a weak episode in 1996, a strong, probably discrete episode, extending from 1998 into 2003. This is followed by a weaker episode beginning in 2004. Thus the ~176 day component in cycle 23 is dominated by the second episode. The second episode is clearly in phase with every second peak of $\Delta(1/R_M^3)$ and therefore in phase with the 176 day first sub-harmonic of the tidal effect of Mercury for the seven years from 1998 to 2003, i.e. over 25 orbits of Mercury.

Figure 5 shows the ~176 day variation during cycle 18 as an example of sub harmonic variation where there are three clearly discrete episodes.

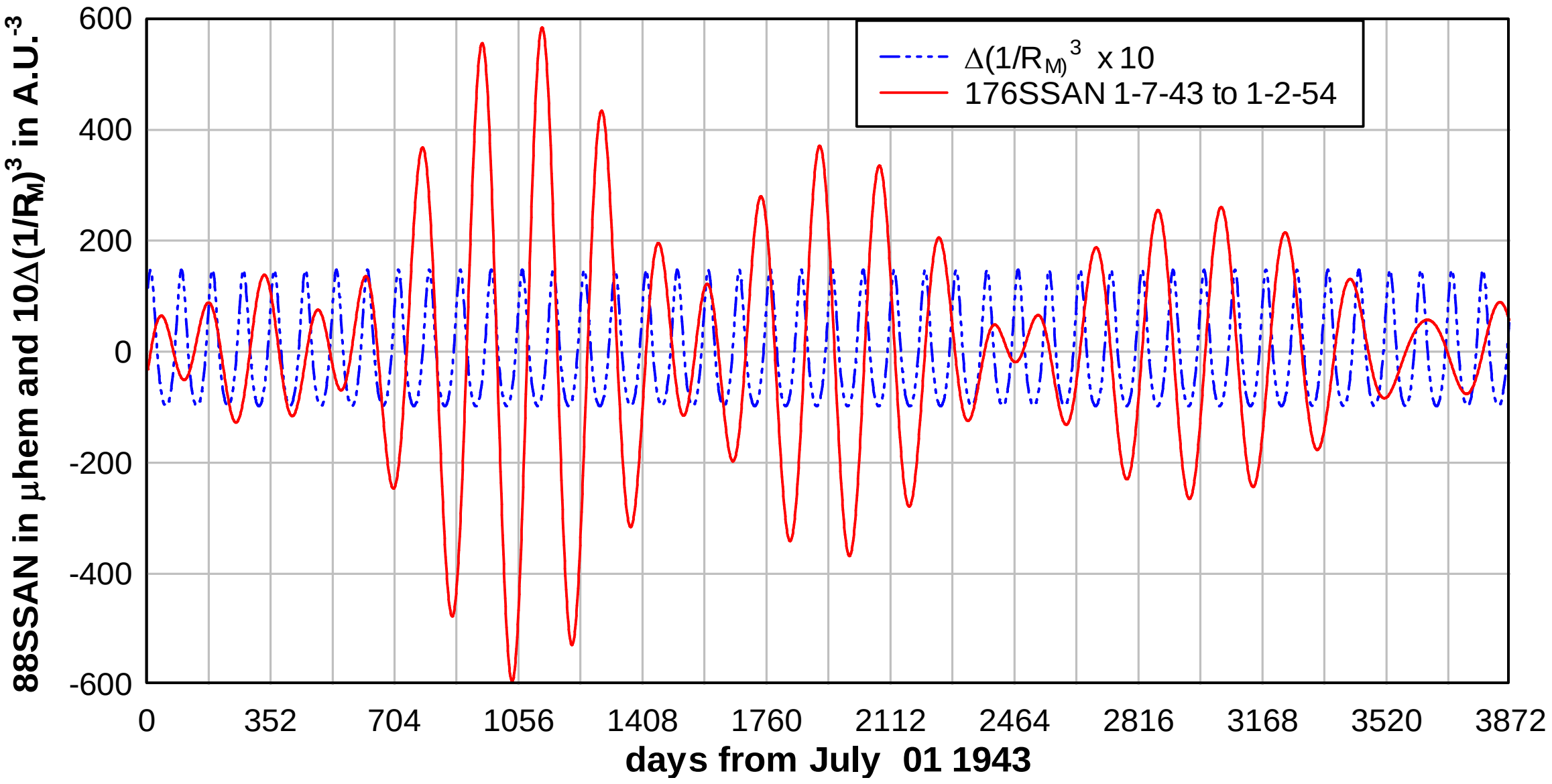

**Figure 5.** The ~176 day period component of SSAN from 1943 to 1954 during solar cycle 18 showing three discrete episodes of sunspot emergence. The time axis is in intervals of 176 days to indicate more clearly the π phase change in the ~176 day period variation between each episode.

Here the peaks of the ~176 day first sub harmonic response in all three episodes occur between sequential pairs of peaks of the Mercury tidal effect. However, with the time axis divided into intervals of 176 days, it is evident that there is a π phase change of the ~ 176 day response between the first and second episode and between the second and third episode. Thus we note that the ~88 day variation and the ~176 day variation, both suffer a π phase change of the variation between episodes if the episodes are discrete.

In a nonlinear system there will always be a stable response at frequencies that are harmonics of the driving frequency. However, sub harmonic response is usually associated with some instability in the system and usually occurs only when some threshold level is exceeded and triggers a system response, Yen (1971). We assume that, in the present case, the instability resulting in sub harmonic response is associated with buoyant instability of magnetic flux, Lou (2000).

In most solar cycles there is considerable overlap between some or all of the episodes. For example Figure 2 shows that in cycle 20 there is strong overlap of three ~88 day episodes. When this is the case it is not possible to discern the clear phase coherence that is obvious in the case of discrete episodes. Further it is difficult to ascertain whether the variation in sequential, overlapping, episodes suffers an exact π phase change between episodes. Reference to Figure 2 shows that solar cycle 14 illustrates a case where there are both discrete and overlapping episodes in the ~ 88 day component of sunspot area North. The variation of the ~88 day component during cycle 14 is compared with $\Delta(1/R_M^3)$ in Figure 6.

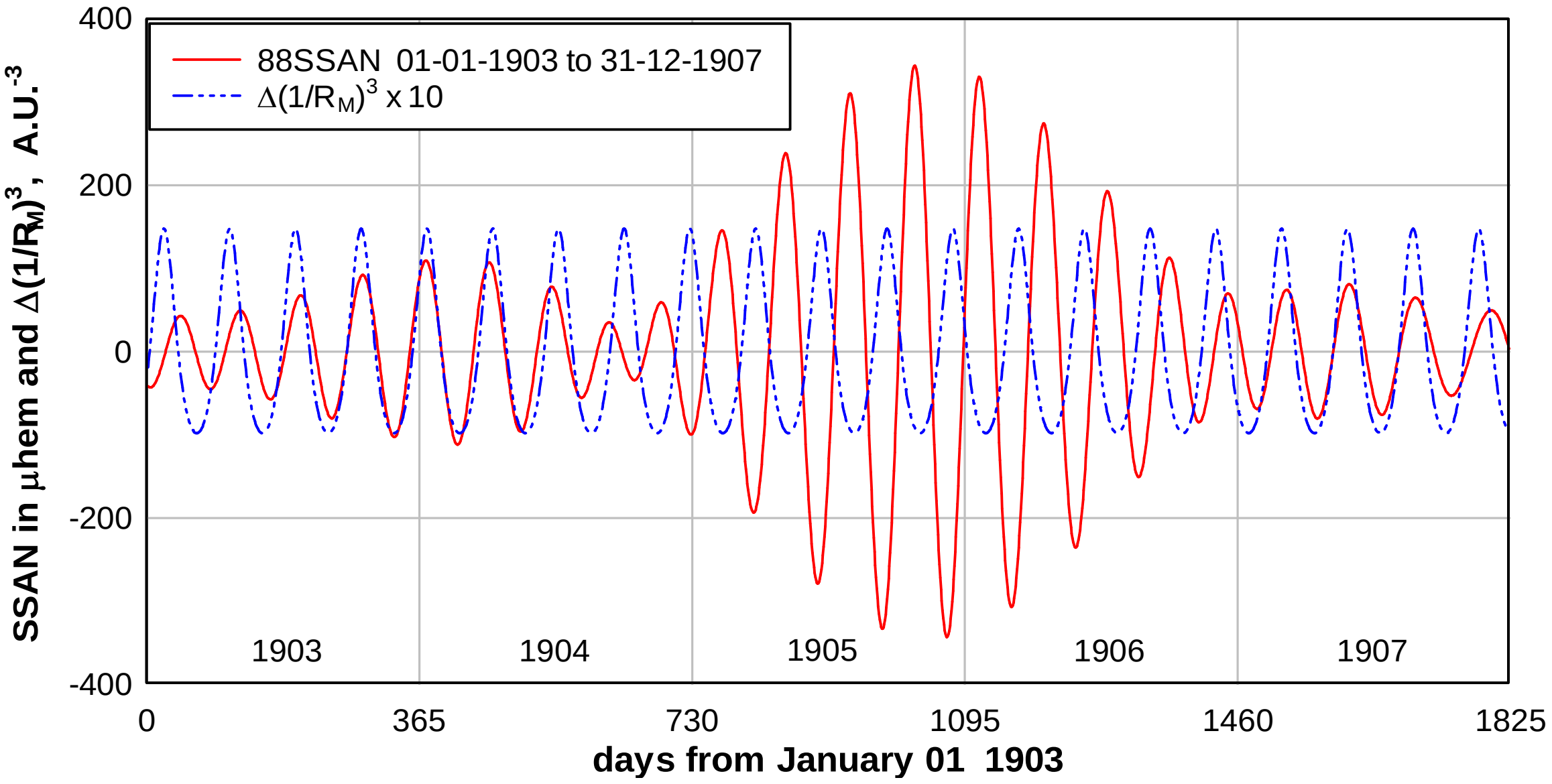

**Figure 6.** The ~88 day component of SSAN for 1903 to 1907 in solar cycle 14 (full line) compared with the variation of $\Delta(1/R_M^3)$. Two weak episodes and one strong episode of sunspot emergence are evident with a $\pi$ phase shift between each episode.

The weak episode in 1903 1904 is discrete from the strong episode in 1905 1906 and a $\pi$ phase change between exact in-phase variation with $\Delta(1/R_M^3)$ to anti-phase variation with $\Delta(1/R_M^3)$ is evident between these two episodes. The variation in 1906 is beginning to overlap the weak in-phase episode in 1907 and it is difficult to discern a simple phase relationship during this interval. It is interesting to note that because the phase of $\Delta(1/R_M^3)$ is known precisely we can infer that, when episodes are discrete, the ~88 day and ~176 day components of sunspot area are phase coherent with the variations in other discrete episodes occurring in the past or future. For example, we know the ~88 day variations in sunspot area between 1903 and 1904 in Figure 6 must be exactly phase coherent with the ~88 day variations in sunspot area between 2002 and 2003, Figure 3, simply because the Mercury tidal effect is a clocklike time reference. It is interesting to note that Lean (1990) investigated the phase self-coherence of the 155 day quasi-periodicity in successive solar cycles using a method of fitting sinusoids to sunspot area data filtered with a band pass filter centred on 155 days. Lean (1990) found that "While these investigations of the 155 day periodicity do indicate approximate phase coherence between adjacent solar cycles, they also illustrate that the 155 day periodicity is not equivalent to a deterministic, sinusoidal oscillation at one fixed period. Rather, the actual period drifts with time, and uncertainties in knowing this period propagate as errors when the cycle is extrapolated too far beyond the time interval in which the actual period has been determined." We will show, in the next section, that the ~155 day quasi-periodicities observed by Lean (1990) and others are actually sidebands of the 176 day first sub harmonic variation in sunspot area due to the Mercury tidal effect and that the sideband periods that occur in a spectrogram obtained over a solar cycle depend on three parameters; (1), the constant planetary periodicity acting as a driver, (2), the variable number of episodes in the cycle and (3), the variable average time interval between episodes in the cycle. The latter two parameters vary from cycle to cycle. Therefore we expect the sideband periods to vary from one solar cycle to the next. We examine the cycle to cycle variation of quasi-periodicities in the next section.

Before leaving this section we briefly examine the phase coherence of the variation of the ~88 day component of sunspot area South (SSAS) with the variation of $\Delta(1/R_M^3)$ during solar cycle 23, Figure 7.

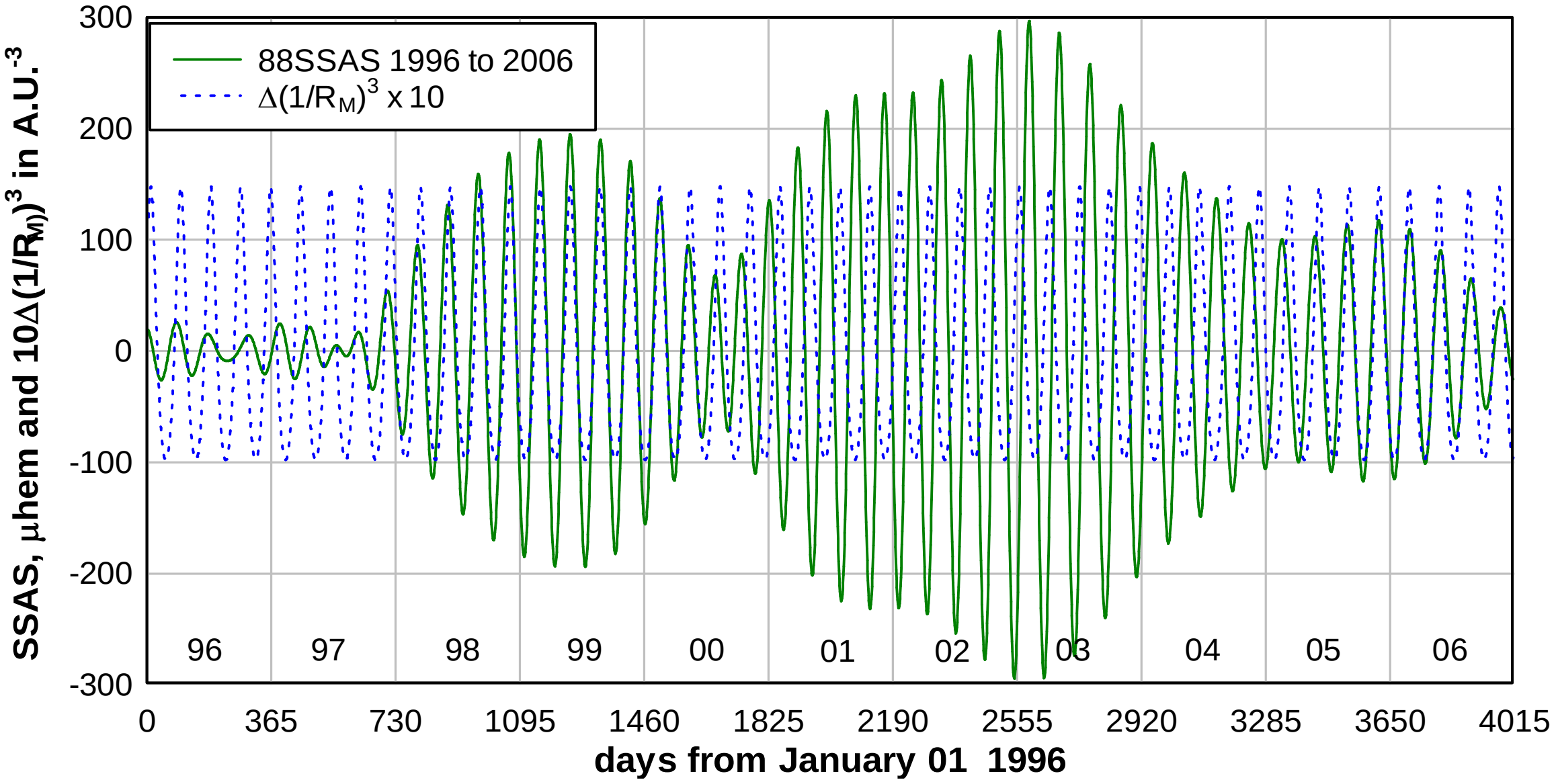

**Figure 7.** The ~88 day component of sunspot area South (SSAS) for 1996 to 2006 in solar cycle 23 showing a strong and discrete episode exactly in-phase with $\Delta(1/R_M^3)$ during 1998 to 2000, followed by a strong predominantly anti-phase component 2001 to 2004 and a weak overlapping exactly in-phase episode 2005 to 2006. There is an indication of a weak overlapping in-phase episode in 2002 2003. Comparison with Figure 3 indicates that the ~88 day components of SSAS and SSAN are varying in phase only in 1998 and 2004.

There is a strong discrete episode during 1998 1999 that is exactly in-phase with $\Delta(1/R_M^3)$. This is followed by a long, apparently single, episode extending over 2001, 2002, 2003, and 2004. The variation in this episode is, on average, in anti-phase with $\Delta(1/R_M^3)$. However, in 2002 2003 the shift is between in-phase and anti-phase and that suggests the presence of an exact in-phase episode occurring between and strongly overlapping two anti-phase episodes. There is a weaker exactly in-phase component in 2005 2006 that overlaps the anti-phase episode ending in 2004 2005. As mentioned earlier when episodes are not discrete but overlap strongly it is difficult to assess the phase relationship. Comparison with the ~88 day SSAN variation, Figure 3, which has several discrete episodes, all exactly in-phase or exactly anti-phase with $\Delta(1/R_M^3)$, shows that the ~88 day components of SSAN and SSAS are both varying exactly in-phase with $\Delta(1/R_M^3)$ only during 1998 and both varying exactly anti-phase with $\Delta(1/R_M^3)$ only during 2004. We find that phase coherence of the ~88 day components of sunspot area in both the northern hemisphere and southern hemisphere of the Sun occurs infrequently. However, when it does occur the phase coherence can be shown to propagate through to either exact in-phase or exact anti-phase variation with $\Delta(1/R_M^3)$ of the ~88 day components of global measures of solar activity such as total sunspot area and sunspot number. However, further examination of sunspot area South and global measures of solar activity is outside the scope of the present article.

**5. Spectral analysis of sunspot area data.**

Since the discovery of quasi-periodicities in solar flare data by Reiger (1984) the discovery and study of quasi-periodicities in a wide range of space variables has been conducted, principally, by means of spectral analysis. Nearly all of the reports of quasi-periodicities in solar related variables referred to in the introduction are based on the discovery of peaks in the 45 day to 500 day period range in spectrograms, periodograms or time-frequency diagrams. Figure 2 showed that sunspot emergence in this period range occurs in episodes lasting from about 1 to 5 years. It follows that any variation in solar activity or solar related variable will be amplitude modulated by this type of episode. The spectrum of an amplitude modulated signal contains sidebands and this introduces an ambiguity in assessing the origin of a particular periodicity. It is clear that before assessing a spectrogram of a variable like sunspot area, in, for example, cycle 23 where sequential discrete episodes occur with $\pi$ phase change between each episode, it is necessary to understand the effect amplitude modulation of this type has on a spectrogram.

**5.1 Spectra of amplitude modulated signals.**

We study the simple amplitude modulation,

$$
\begin{aligned}
y &= [A+\sin(2\pi t/730)]\sin(2\pi t/88) \\
&= [A+\sin(2\pi f_m t)]\sin(2\pi f_1 t) \\
&= A\sin(2\pi f_1 t) - 0.5\{\cos(2\pi(f_1 + f_m)t) - \cos(2\pi(f_1 - f_m)t)\} \qquad (3)
\end{aligned}
$$

The term in square brackets is a two year, 730 day, modulation that modulates the 88 day sinusoid of frequency $f_1$ = 1/88 = 0.01137 days$^{-1}$, Figures 8A and 9A. The modulation period, $T_m = 1/f_m$ = 730 days, is the time interval between the maxima of episodes having variations with the same phase. When A >> 1 there is no modulation, no episodes, and a spectrogram of y contains a single peak at 88 days. When A ~ 1 there is strong amplitude modulation and side bands appear at $f_S = f_1 +/- f_m$ in the spectrogram, see Figures 8A and 8B. When A << 1 the modulation is said to “cross zero”, the sign of the 88 day sinusoid is reversed when the modulating term becomes negative, as a result there is a $\pi$ phase shift in the 88 day signal between one episode and the next, and nearly all signal power is shifted from the central peak at $f_1$ to the two sideband peaks at $f_1 +/- f_m$, Figures 9A and 9B. The time axis of the Figures is divided into 88 day intervals so that phase changes can be followed. The time variations in Figure 8A and 9A are shown over 1056 days for clarity. However the spectrograms or FFTs in Figures 8B and 9B are calculated over 3650 days to provide better frequency resolution.

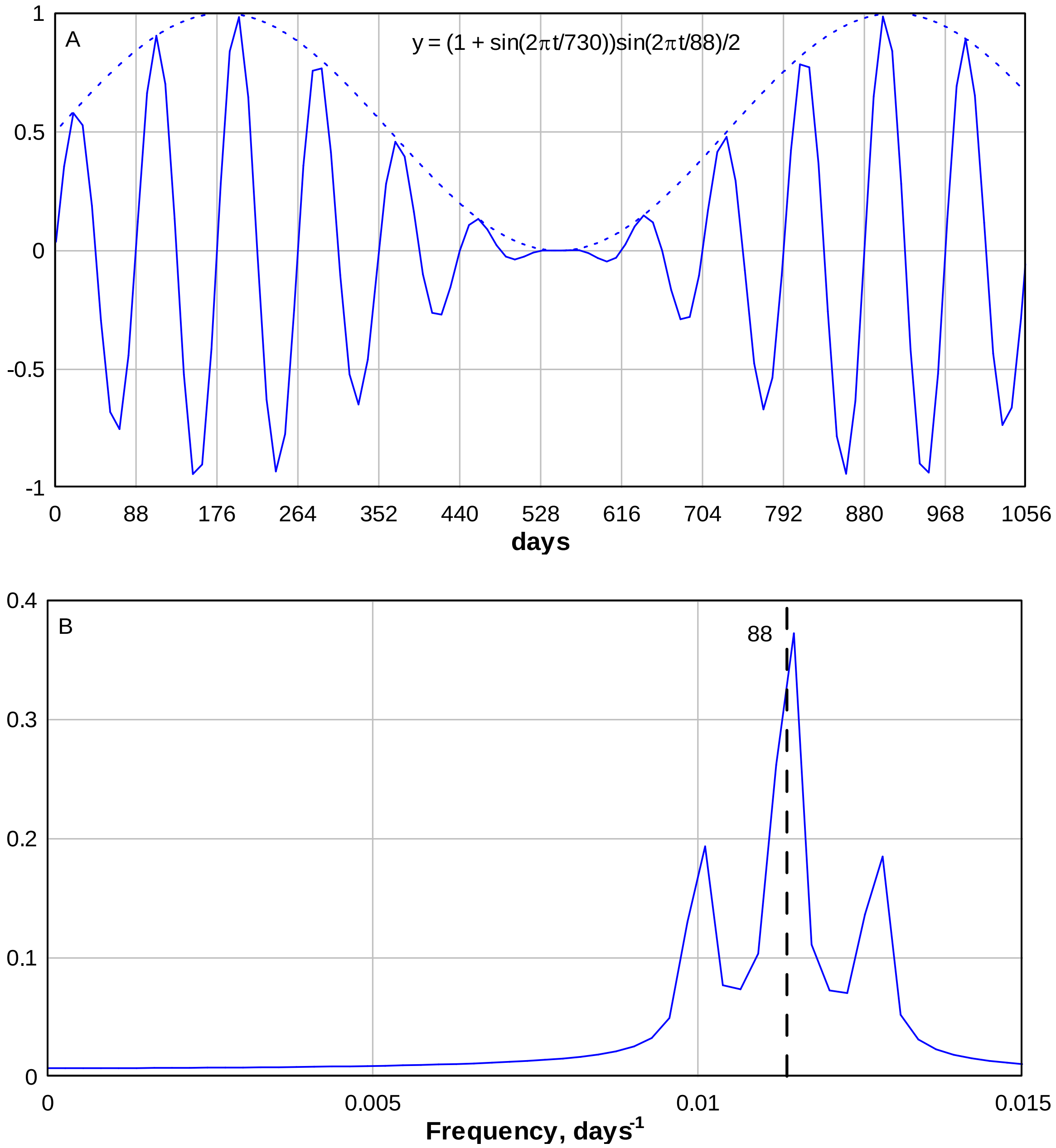

**Figure 8.** The variation (A) and the spectrogram of the variation (B) of an amplitude modulated sinusoid when A = 1 in equation 3. The time axis of the variation is in intervals of 88 days so that any phase change can be followed. In this case there is no phase change from one modulation maximum to the next.

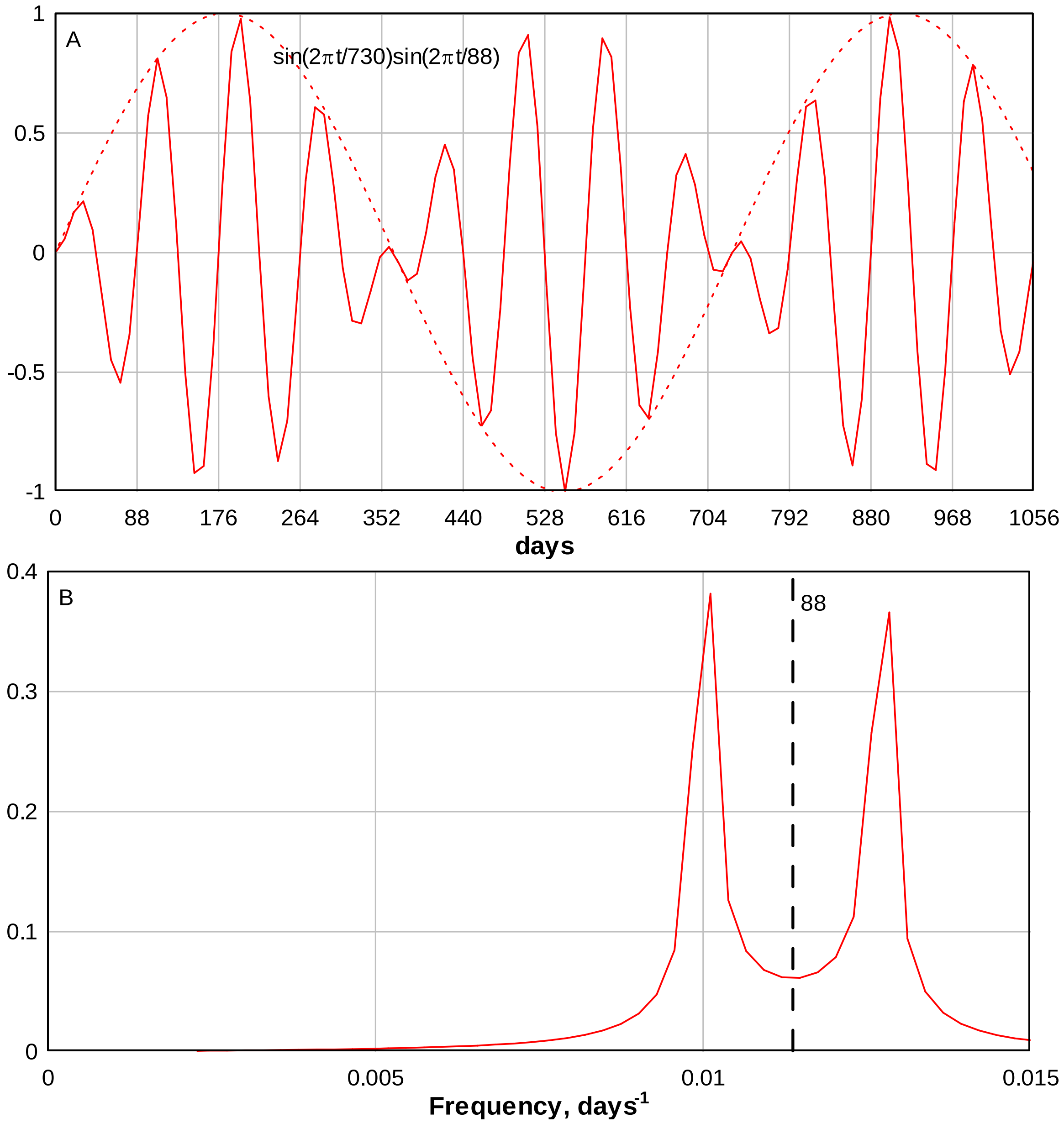

**Figure 9.** The variation (A) and the spectrogram of the variation (B) of an amplitude modulated sinusoid when A = 0 in equation 3. The time axis of the variation is in intervals of 88 days so that any phase change can be followed. In this case there is a $\pi$ phase change from one modulation maximum to the next.

It is evident from the simulations above that the variation during any one episode is at 88 day period and a spectrogram made over one episode will have the major peak at 88 day period. However, if a spectrogram is made over several episodes when A ~ 0 the spectrogram may show no peak at 88 days and show only sidebands on either side of a central minimum centred on 88 days as in Figure 9B.

**5.2 Interpreting spectrograms of sunspot area in individual solar cycles.**

The simple modulation relation of equation 3 is a useful means of interpreting the effect that the episodes of sunspot area emergence observed in Figures 2 to 7 have on the corresponding spectra. For example, in the ~88 day component of SSAN data for solar cycle 23, Figure 3, four discrete episodes occur and there is a π phase shift between the variations in succeeding episodes, so the spectrum can be interpreted by equation 3 with A ~ 0. The episodes are separated, on average, by ~1.5 years or ~ 550 days. Thus the modulation period, the time interval between one episode and the next episode of the same phase, is $T_m$ = 1100 days. Thus, A = 0, $T_1$ = 88 days, $T_m$ = 1100 days in equation 3 would simulate this case. The expected spectrogram should evidence a minimum at $T_1$ = 88 days or $f_1$ = 0.0114 days$^{-1}$ and strong sidebands at $f_1$ -/+ $f_m$ i.e. at frequencies 0.0104 days$^{-1}$ and 0.0123 days$^{-1}$. The corresponding sideband periods are 81 days and 96 days. The observed spectrogram of the unfiltered SSAN data for the eleven years between 1996 and 2006 is shown as the blue dotted line in Figure 10. The spectrograms of the ~ 88 day variation and the ~ 176 day variation are also shown in Figure 10.

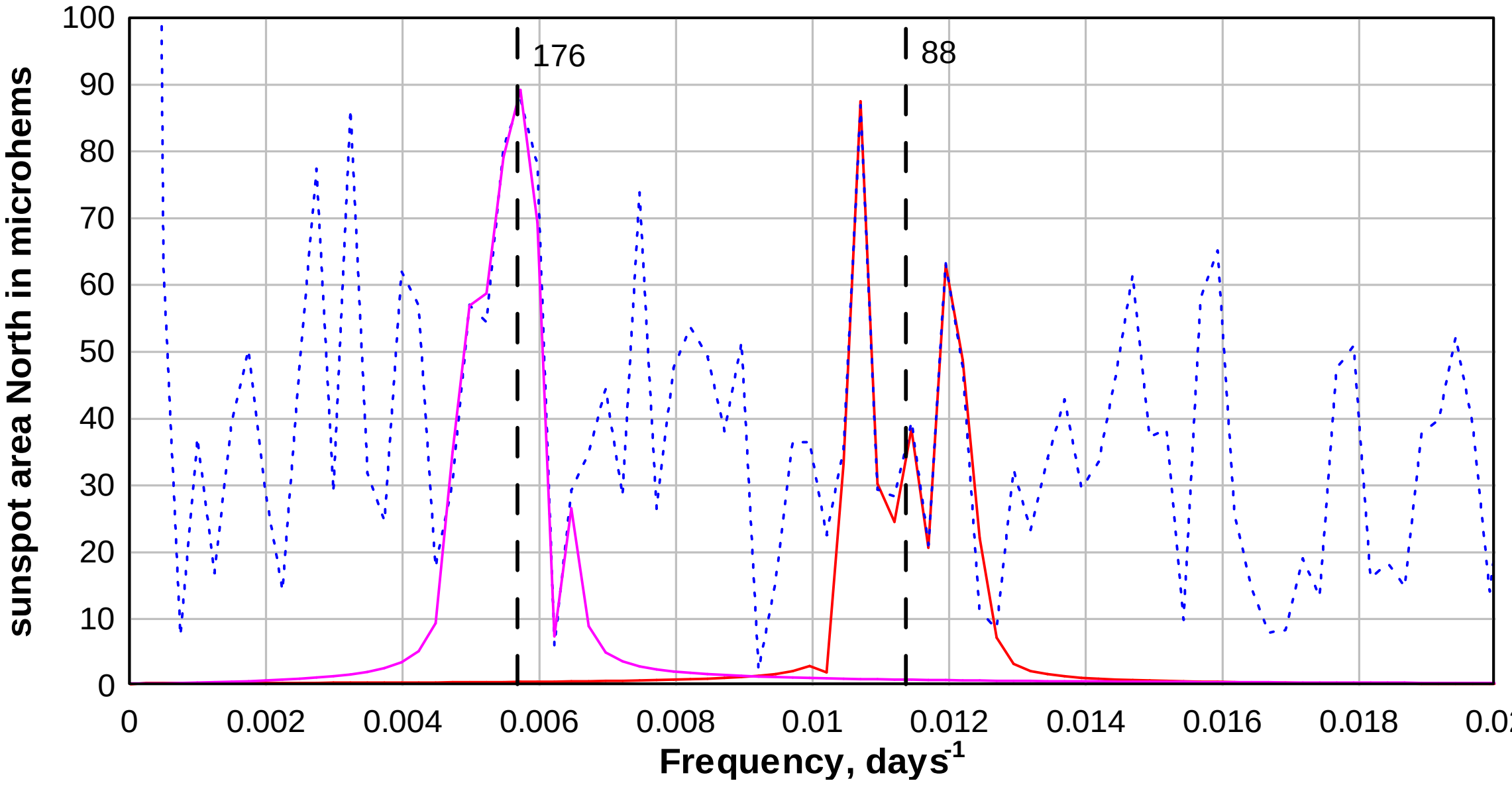

**Figure 10.** The spectrogram of daily SSAN data during the interval 1996 to 2006 in solar cycle 23, (blue dots). The spectrograms of the ~ 88 day and ~ 176 day components of SSAN are superimposed for comparison.

The spectrum around the 88 day period shows the two sidebands on either side of a minimum at 88 day period as expected when the episode modulation crosses zero and the 88 day period variations in successive episodes differ by π in phase. As indicated above this situation is expected for A ~ 0 in equation 3. The spectrum also shows a strong peak at 176 days. From the simulations outlined above this is characteristic of relatively weak modulation i.e. A > 1 in equation 3. We know from Figure 2 and Figure 4 that there is a long, strong, episode of the ~176 day component of sunspot area North in solar cycle 23. This is followed by a short, weak episode. Thus the episodes of 176 day variation are dominated by the effect of the strong episode and a spectrogram obtained over cycle 23 should approximate the type of spectrogram obtained over a single episode i.e. the spectrogram should exhibit a strong peak at 176 days with relatively weak sidebands. Thus the spectrograms of Figure 10 are consistent with the type of episodes observed in cycle 23 for the ~176 day component as well as for the type of episodes observed in cycle 23 for the ~88 day component of the SSAN data. We note that, without knowledge of the occurrence of episodes of ~88 day and ~176 day variation in sunspot emergence, the evidence of the in-phase or and anti-phase coherence of the variation within each episode with $\Delta(1/R_M^3)$, and interpretation with equation 3, it would be difficult to discern any effect of 88 day periodicity in the spectra because the peak at 88 days in Figure 10 is insignificant.

An example of a solar cycle that is dominated by one strong episode of the ~176 day component and one strong episode of the ~88 day component is, from Figure 2, solar cycle 22. In this cycle there is a dominant ~88 day episode and a dominant ~176 day episode and we would, therefore, expect a spectrogram of sunspot area to show a strong peak at ~88 day period and a strong peak at ~176 day period and relatively weak sidebands. Using the standard FFT method we find that this is the case although we do not reproduce the spectrum here. However, in support of this interpretation, we note that Oliver and Ballester (1995), using the more powerful spectral analysis methods of Lomb-Scargle periodogram and the Maximum Entropy Method, found the most significant peak in sunspot area spectrum during cycle 22 was at 133 nHz (87 days). A significant peak at 66 nHz (175 days) was also found in most of the intervals in cycle 22 investigated by Oliver and Ballester (1995).

We chose solar cycle 23 to illustrate the effect episodic modulation has on the spectra of sunspot activity because the episodes of the ~88 day component during solar cycle 23 were discrete and relatively simple to interpret in terms of equation 3. In solar cycles where the episodes overlap and/or are variable in strength e.g. cycle 19 and 20 as shown in Figure 2, simulations more detailed than the simulation of equation 3 would need to be applied.

The main objective of this section was to demonstrate the potential for drawing ambiguous conclusions from spectrograms when attempting to discover or study quasi-periodicities in solar activity related data. It is clear from the above simulations and observations that a sideband of a modulated signal may appear as a significant peak in a spectrogram when a peak at the period of the driving mechanism is insignificant or absent. For example, if one wishes to discover ~160 day "quasi-periodicities" in sunspot area data one can note that the required sideband frequency, $f_S$, is 0.00625 $days^{-1}$. Use of the sideband relation $f_S = f_2 +/- f_m$ with $f_2$ = 1/176 $days^{-1}$ indicates that the required modulation frequency is $f_m$ = 0.00057 $days^{-1}$ or modulation at a period $T_m$ = 4.8 years. Thus intervals with two or more discrete episodes of the ~176 day component where the episodes are separated by about 2.4 years will be intervals when ~160 day "quasi-periodicities" are likely to be discovered. From Figure 2 we can see that solar cycle 18 is a good example to illustrate this effect as it has three discrete episodes of ~176 day variation each separated by about 2.4 years.

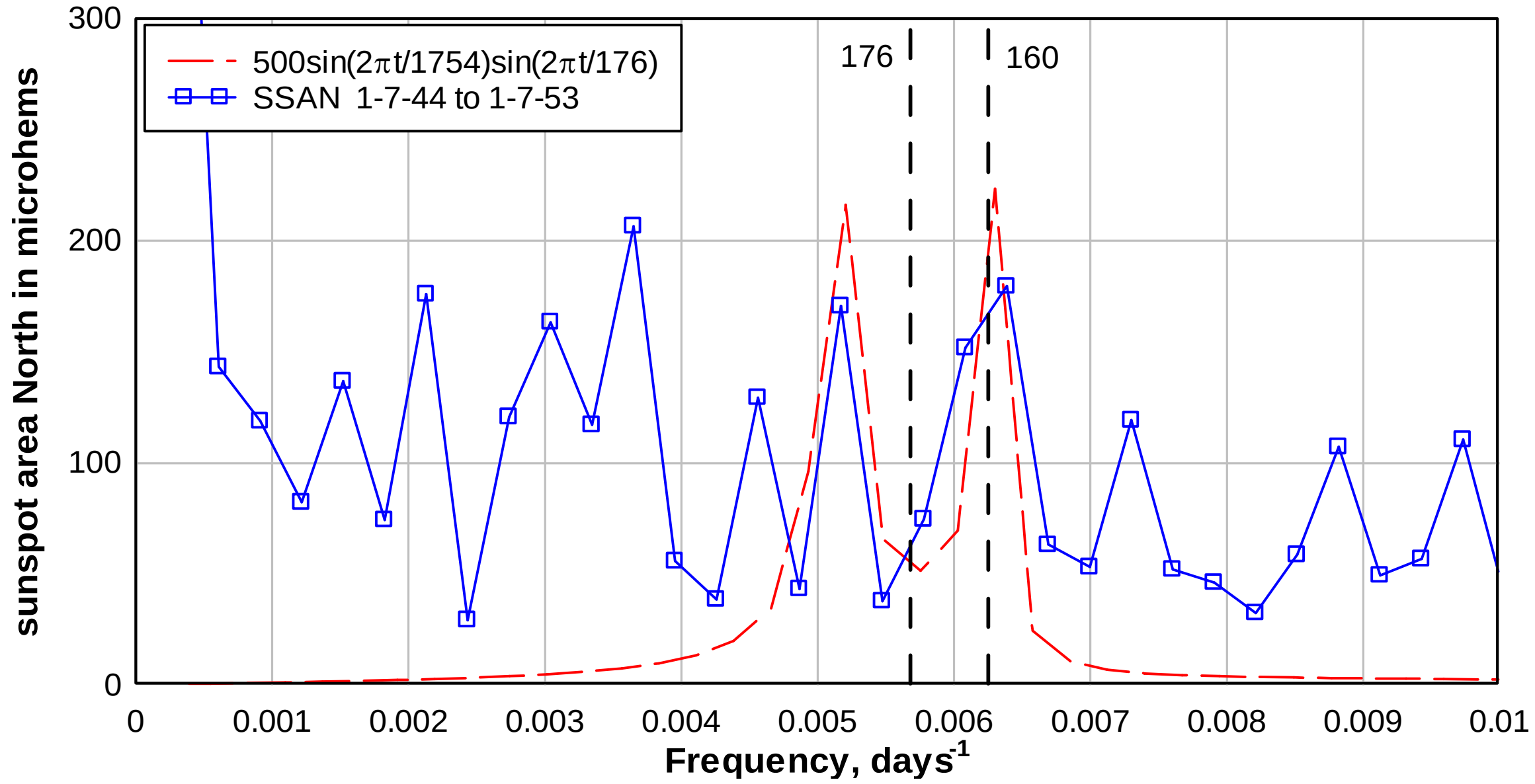

**Figure 11.** Illustrates the result of the method for "discovering" ~ 160 day periodicities in sunspot area data. The broken red line shows the spectrogram due to the variation of equation 3 with A = 0, $f_m$ = 0.00057 $days^{-1}$ and $f_1$ = 1/176 $days^{-1}$ and the blue squares show the observed spectrogram between 1944 and 1953 in solar cycle 18.

In Figure 11 the broken red line shows the spectrogram of a simulated variation obtained from equation 3 with when applying the "discovery" procedure outlined above with $T_2$ = 176 days and $T_m$ = 4.8 years. The full blue line is the actual spectrogram of the SSAN data observed during solar cycle 18, in the interval July 1 1944 to July 1 1953. Evidently a ~160 day quasi-periodicity has been "discovered" in SSAN data during this interval. However, the actual driving periodicity is the 176 day sub harmonic of the radial variation of Mercury. We note again that the period of a sideband peak depends on three parameters; the constant orbital period of Mercury or periods of its sub harmonics, the varying number of episodes in the interval considered and the average time interval between episodes of the same phase.

The results of Section 4 indicated that a π phase shift occurs between sequential discrete episodes of filtered data. This Section demonstrated that spectrograms taken over several discrete episodes during a solar cycle have a minimum at the driving period with sidebands on either side, the periods of which depend on the time interval between episodes. We infer from this result that it may sometimes be more useful to search for deep minima rather than high peaks in the spectrograms of variables connected with solar activity. We also infer that some spectrograms obtained in previous research could be interpreted from this point of view. A good example is the spectrum of daily counts for flares in cycles 20 and 21, Bogart and Bai (1985), Figure 2C, where a deep minimum at 176 days (0.065 μHz) has several closely spaced sidebands on either side of the minimum.

This leads us to consider the overlap of the spectra associated with the higher sub harmonic periods of Mercury. We have seen, in this section and in the previous section that discrete episodes, when these occur, tend to occur at intervals of 1.5 – 3.5 years.

Thus the modulation period, $T_m$, varies between 3 – 7 years and the sidebands in the spectra due to episode modulation tend to be spaced from the sub harmonic frequency by about 0.0004 days$^{-1}$ to about 0.0009 days$^{-1}$. Thus when considering spectra associated with higher sub harmonics of Mercury, for example, the third, 246 day, sub harmonic and the fourth, 352 day, sub harmonic at, respectively, 0.00406 days$^{-1}$ and 0.00284 days$^{-1}$ frequency we can expect overlap of the individual spectra to occur. An illustration of the type of overlap expected for the case when $T_m$ = 4 years, is shown in Figure 12.

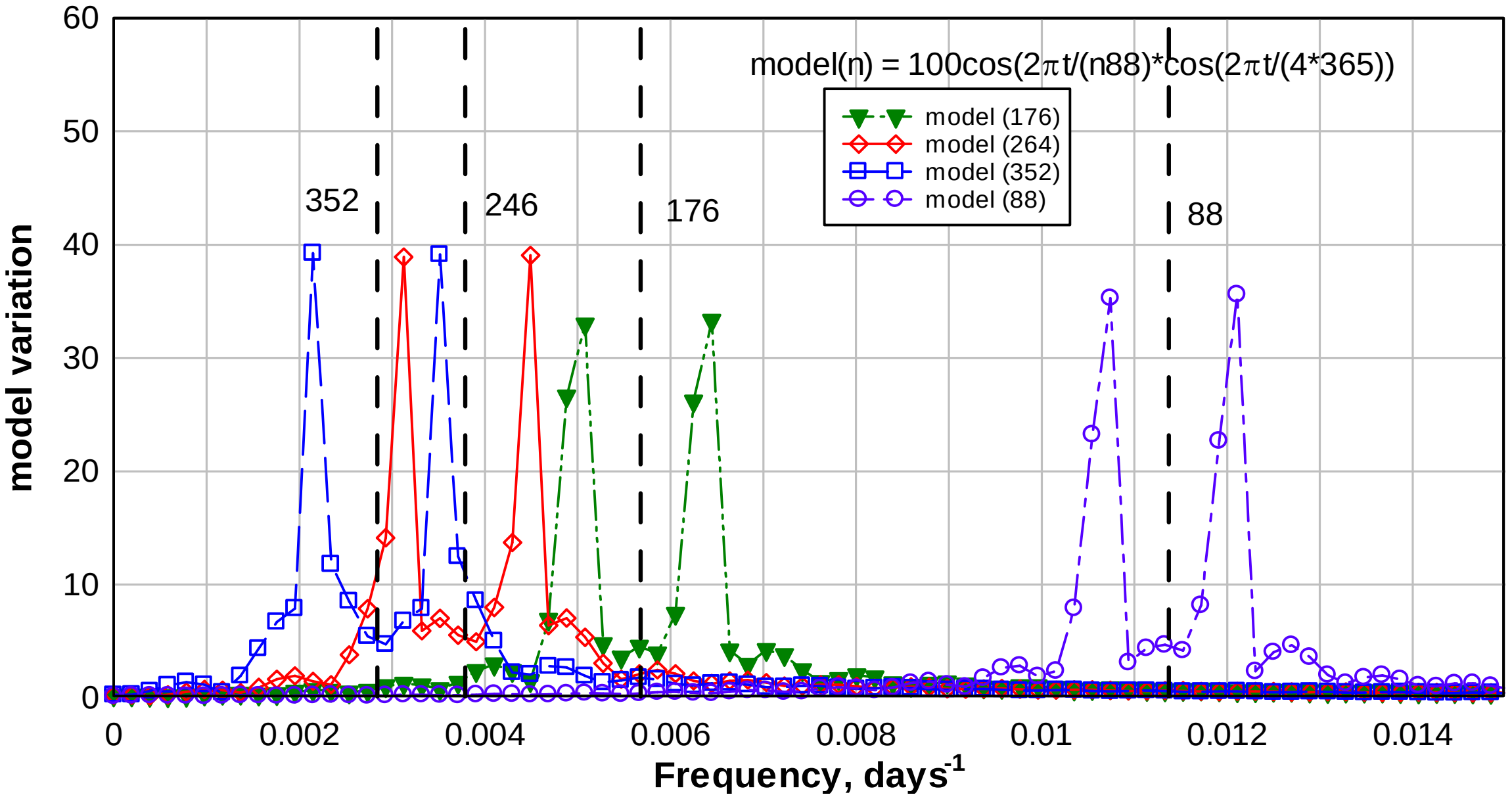

**Figure 12.** An illustration of the overlap of the sidebands in spectra associated with higher number sub harmonics of the tidal effect of Mercury. For sub harmonics higher than the 176 day sub harmonic we expect considerable overlap of the individual spectra.

It is evident that in spectra taken over just a single solar cycle, approximately 4000 days, the spectral resolution is so low that identification of individual peaks with the sidebands associated with the higher sub harmonic spectra would become problematic. We therefore turn attention to spectral analysis of records extending over several solar cycles when higher resolution is available.

### 5.3 Frequency spectra obtained over the entire sunspot area record.

It is evident from the time variation of the 88 day and 176 day components shown Figure 2 that when assessing spectra obtained over multiple solar cycles the modulation due to the ~11 year solar cycle variation in sunspot area, must be taken into account. At the start of a solar cycle the observed sunspot area rises from near zero to a maximum then falls to near zero at the end of the cycle. Thus the amplitude modulation due to a solar cycle can, as a first approximation, be expressed as $1 + \cos(2\pi t/(11*365))$ or as $1 + \cos(2\pi f_n t)$ where $f_n = 1/4015 = 0.000249$ days$^{-1}$. Thus a model of the variation of the 176 day component of sunspot area over several cycles can be expressed, in terms of periods, by

$$model = \cos(2\pi t/176)[A + \cos(2\pi t/T_m)](1 + \cos(2\pi t/(11*365))$$

or, in terms of frequencies, by

model = $\cos(2\pi t f_2 t)[A + \cos(2\pi t f_m t)](1 + \cos(2\pi f_n t))$ (4)

where $f_2 = 1/176 = 0.00568$ days$^{-1}$, $f_m = 1/T_m$ days$^{-1}$, and $f_n = 1/4015 = 0.000249$ days$^{-1}$. With A = 0, i.e. strong episodic modulation, on expanding equation (4) we find

model = $\cos(2\pi(f_2 +/- f_m)t) + 0.5\cos(2\pi(f_2 +/- f_m + f_n)t) + 0.5\cos(2\pi(f_2 +/- f_m - f_n)t)$ (5)

When A is non zero there are three additional terms:

$A\cos(2\pi f_2 t) + (A/2)[\cos(2\pi t(f_2 + f_n)t) + \cos(2\pi(f_2 - f_n)t)]$ (6)

Thus, with A = 0 the model variation has six frequency components, when A > 0, nine components. The corresponding frequencies in days$^{-1}$ and the periods in days are tabulated in Table 1 for the case when the modulation period of episodes is $T_m$ = 7 years, $f_m = 1/(7*365) = 0.000391$ days$^{-1}$ and A = 0.

Table 1. The frequencies and periods in a model spectrum when A = 0, $f_2 = 1/176$ days$^{-1}$, $f_m = 1/(7*365)$ days$^{-1}$ and $f_n = 1/(11*365$ days). Frequencies are in days$^{-1}$ and periods in days.

| | | | | | |
|---|---|---|---|---|---|
| $f_2+f_m$ | 0.006073 | 164.6 | $f_2-f_m$ | 0.005291 | 189.0 |
| $f_2+f_m+f_n$ | 0.006322 | 158.2 | $f_2-f_m+f_n$ | 0.005540 | 180.5 |
| $f_2+f_m-f_n$ | 0.005824 | 171.7 | $f_2-f_m-f_n$ | 0.005042 | 198.3 |

The FFT of the model variation in this case is shown in Figure 13.

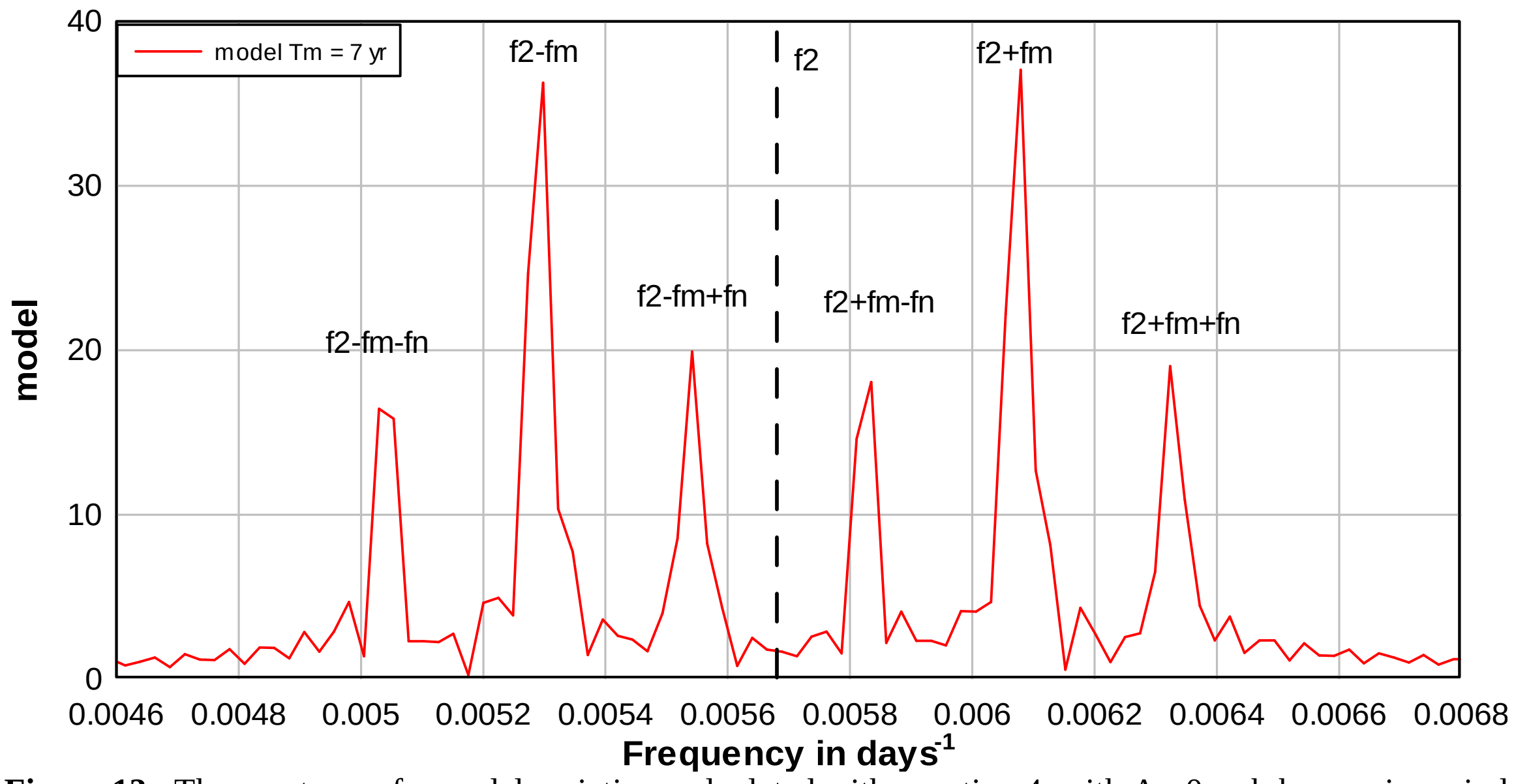

**Figure 13.** The spectrum of a model variation, calculated with equation 4, with A= 0, sub harmonic period $T_i$ = 176 days, episode modulation period $T_m$ = 7 years and solar cycle modulation period $T_n$ = 11 years.

We notice that there is no peak at the sub-harmonic frequency, $f_2$, and therefore no component in the model at 176 day period. This is a result of the $\pi$ phase shift between one discrete episode and the next as modelled by equation 4. All six of the sideband frequencies depend on $f_m$, i.e. on the time interval, $T_m$, between discrete episodes of the 176 day component with the same phase relationship, either in-phase or anti-phase, to the 176 day variation. Assuming the solar cycle frequency, $f_n$, is constant, if $f_m$ also remained constant, the variation of each component of the model would be closely locked to the variation of the other five components and also closely locked to the variation at the sub-harmonic frequency, $f_2$, associated with the first sub harmonic of the variation of the orbital radius of Mercury. As an illustration of this close locking, consider the variation of a sinusoid at the sub harmonic period 176 days and the variation of the model component with period 158.2 days, see Table 1. As 158.2 days is 0.899 ~ 9/10 of 176 days it follows that the two variations will come back into nearly the same phase relationship after nine 176 day cycles and ten 158.2 period cycles. An illustration of this closely locked relationship is shown in Figure 14.

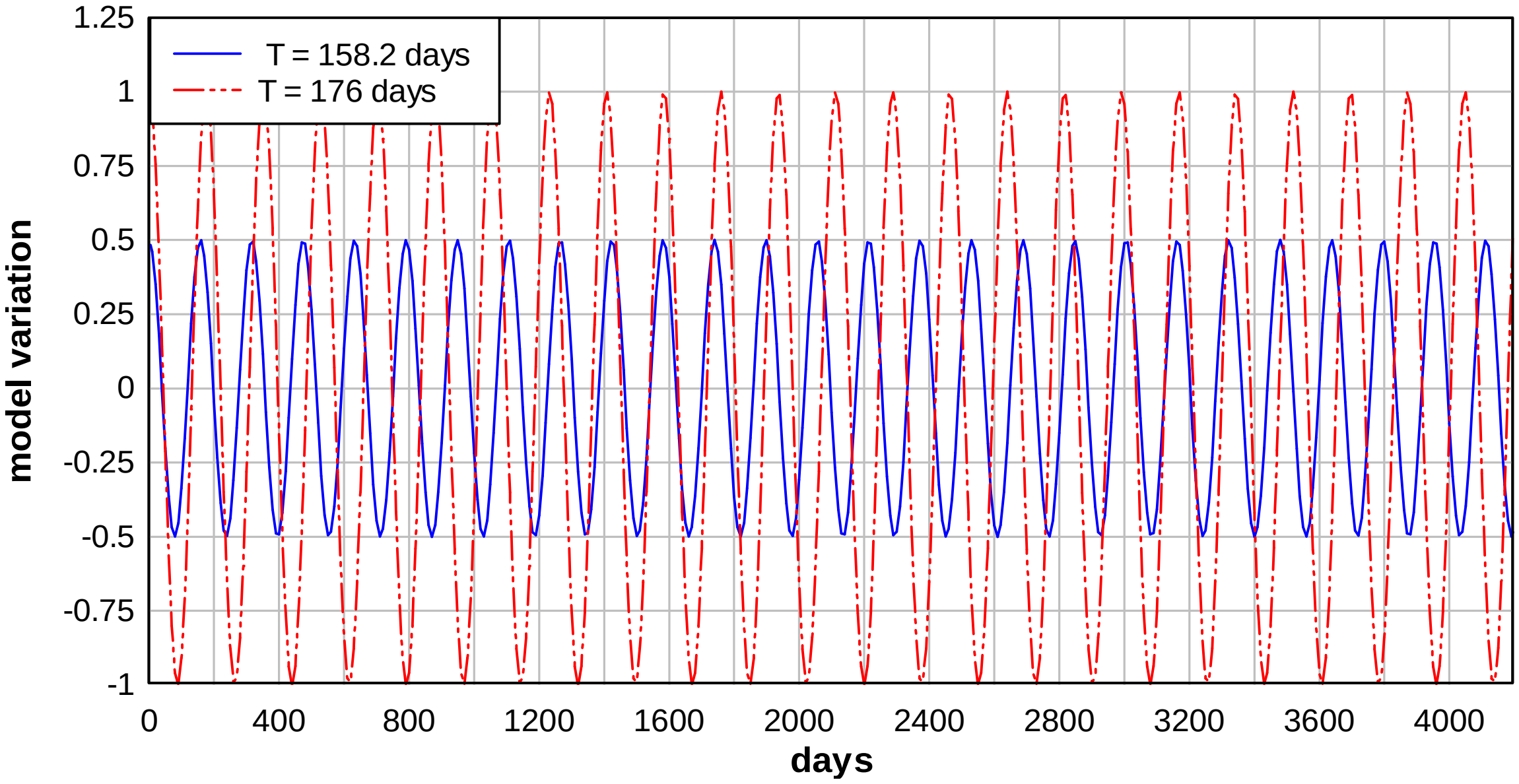

**Figure 14.** An illustration of how a variation at the Reiger sideband period, 158.2 days, returns to be closely in-phase with a variation at the sub harmonic period, 176 days, after nine sub harmonic cycles and ten Reiger cycles indicative of close locking between the two cycles.

If we identify the variation at 158.2 days with the much studied Reiger quasi-periodicity and the 176 day variation with the first sub harmonic of the Mercury tidal variation we can conclude as follows. Although the Reiger variation is not phase coherent with the Mercury tidal variation it is closely locked to the Mercury variation in the sense of returning to a near in-phase condition after a fixed number of Reiger or Mercury cycles, provided $f_m$ and $f_n$ remain constant. Because the Mercury variation is a precise clock it follows that, provided the episode modulation frequency, $f_m$, and the solar cycle frequency, $f_n$, remain constant, the Reiger variation will be nearly self-coherent from solar cycle to solar cycle. If $f_m$ and $f_n$ are not constant the Reiger periodicity will change from one solar cycle to the next. For example, if $T_m$ changes from 7 years in one solar cycle to 6 years in the next the Reiger periodicity changes from 158.2 days to 156.5 days, a difference of 1.7 days, so that at the end of a second 11 year long solar cycle a maximum in the Reiger variation will have shifted by 43 days i.e. by about one quarter of the period of a cycle of the Reiger periodicity. As we expect the shifts from solar cycle to solar cycle to be randomly distributed between positive or negative shifts we expect the average shift after several solar cycles to accumulate as a random walk i.e. to increase as the square root of the number of solar cycles. Extending the previous example, after four solar cycles where $T_m$ varied randomly between 6 years and 7 years we would expect the phase shift to be, on average, 86 days and the phase of the Reiger periodicity to be shifted relative to the phase of the Reiger periodicity in the first solar cycle by, on average, $(86/158)2\pi \sim \pi$ radians and to be, on average, out of phase with the Reiger variation in the first solar cycle. Therefore the model outlined above is consistent with the observations of Lean (1990) of approximate phase self-coherence of the Reiger periodicity in successive solar cycles and the observation, also by Lean (1990), that the phase self-coherence of the Reiger periodicity becomes progressively poorer as the number of successive solar cycles examined for phase self-coherency increases.

While the model outlined above, equation 5, explains the observed phase self-coherence of Reiger periodicity in sunspot area data it also explains the observed variability of quasi-periodicities from solar cycle to solar cycle. This raises the question of whether it is feasible to obtain meaningful results from the examination of frequency spectra or periodograms obtained from sunspot area data extending over several solar cycles, e.g. Lean and Bruekner (1989), or over the entire sunspot area record, e.g. Lean (1990). The problem is illustrated in Figure 15 where the spectrum of SSAN data for the entire record 1876 to 2015 is compared with model spectra obtained when the episode modulation period $T_m$, is varied progressively from 4 to 8 years while the solar cycle period, $T_n$, is held constant at 11 years.

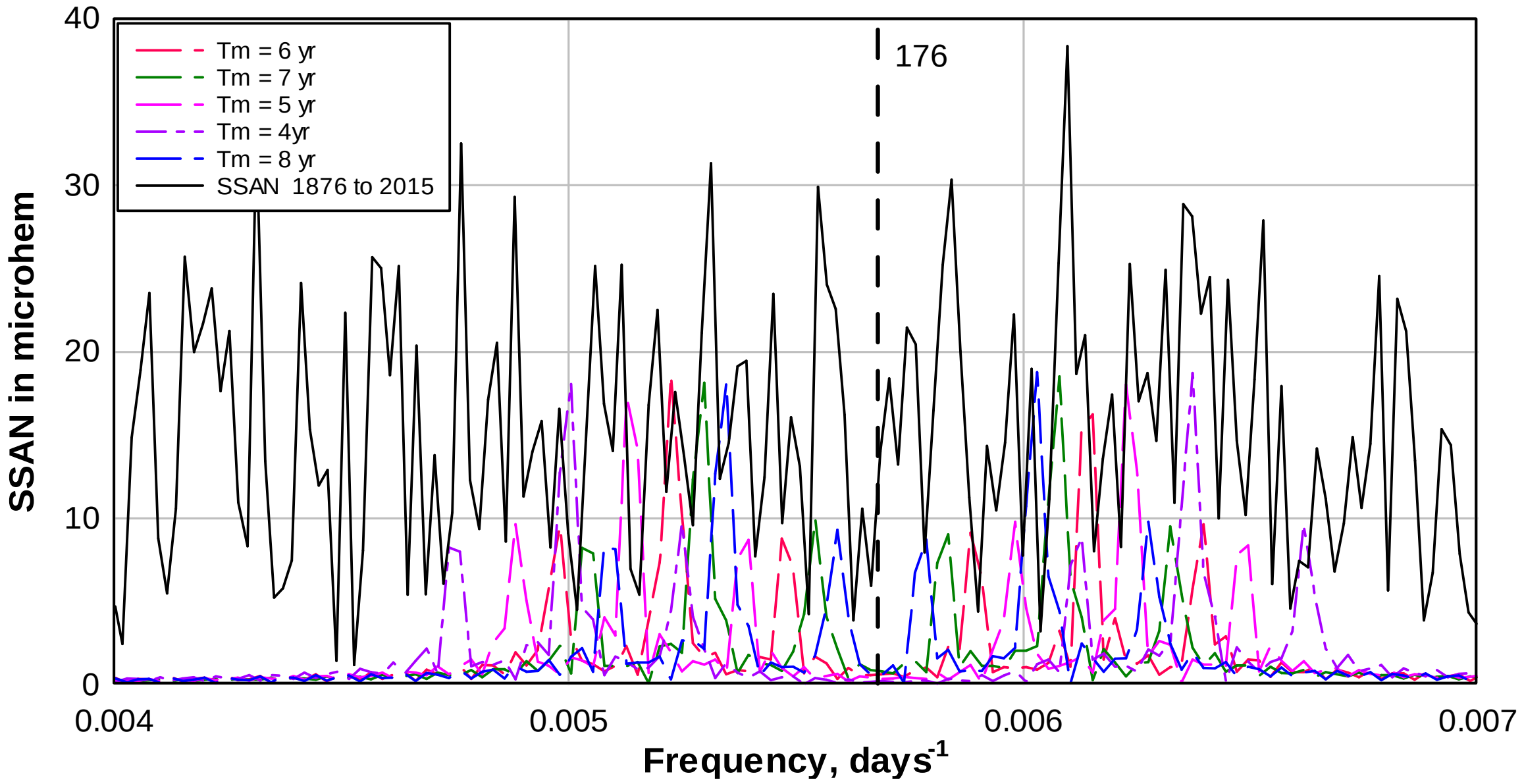

**Figure 15.** Individual model spectra associated with the sub harmonic at 176 days when the episode modulation period $T_m$ is varied from 4 to 8 years. Also, shown for comparison, the spectrum of the entire SSAN data 1876 to 2015.

It is evident from Figure 15 that a broad range of sideband peaks can be expected on either side the Mercury sub-harmonic period. Thus identifying any average model that provides a fit to the observed spectrum is a challenge. For example, by comparing the spectrum of a model variation with $T_m$ = 7 years with the spectrum obtained from the entire SSAN data, Figure 16, we can discern some peak correspondence. However, demonstrating that this correspondence is significant is difficult. We show below that, using spectrum averaging methods similar to those employed by Lean (1990), noise in the spectrum can be reduced and an average model fit more accurately discerned.

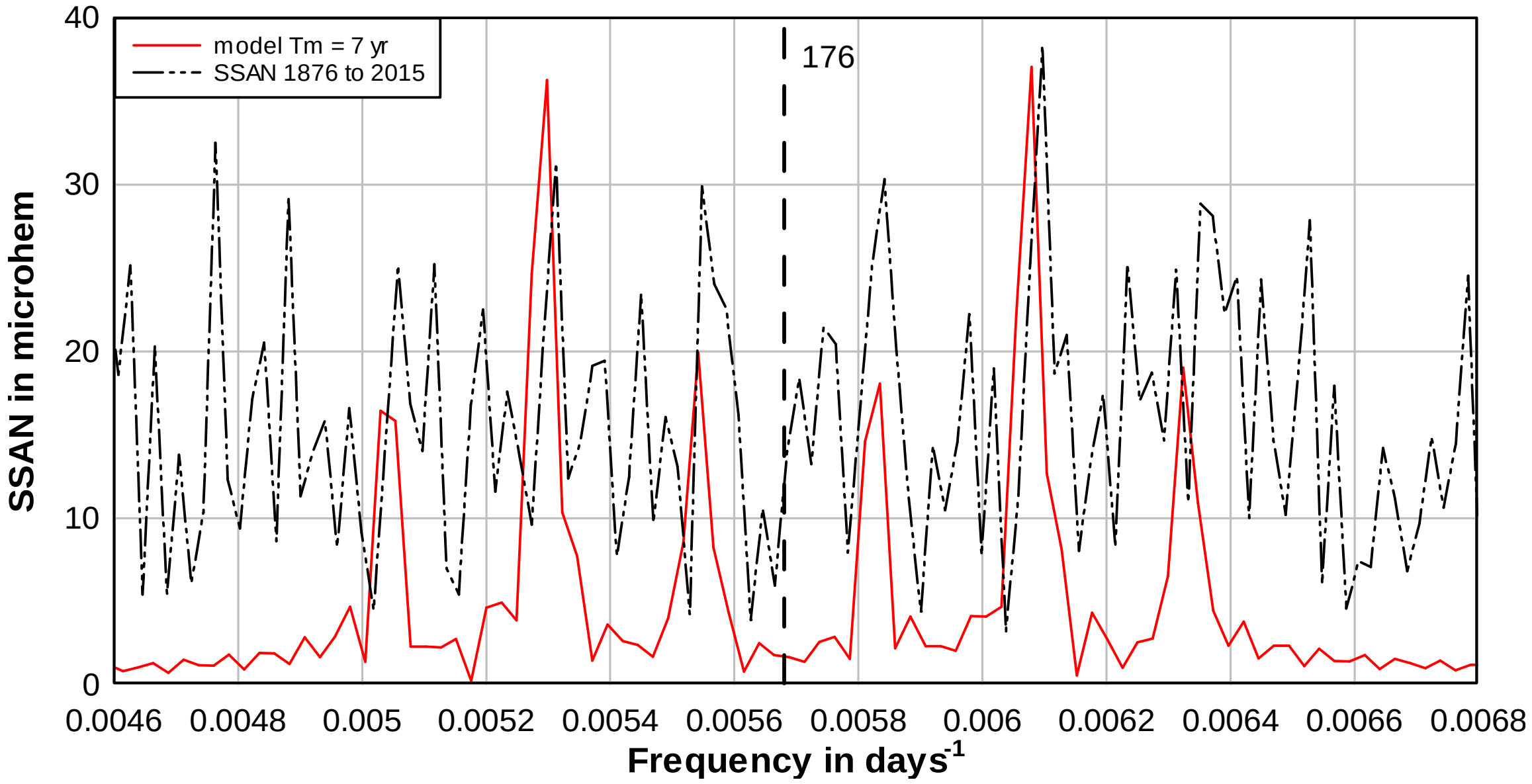

**Figure 16.** The model spectrum associated with the sub harmonic at 176 days when A = 0, the episode modulation period $T_m$ is 7 years and the solar cycle modulation period is 11 years. Also, shown for comparison, the spectrum of the entire SSAN data 1876 to 2015.

To establish the stability of the SSAN spectrum we obtain spectra from the first half and from the second half of the entire SSAN record, 1876 – 2015. Figures 17 and 18 show the two spectra as well as model fits obtained by eye.

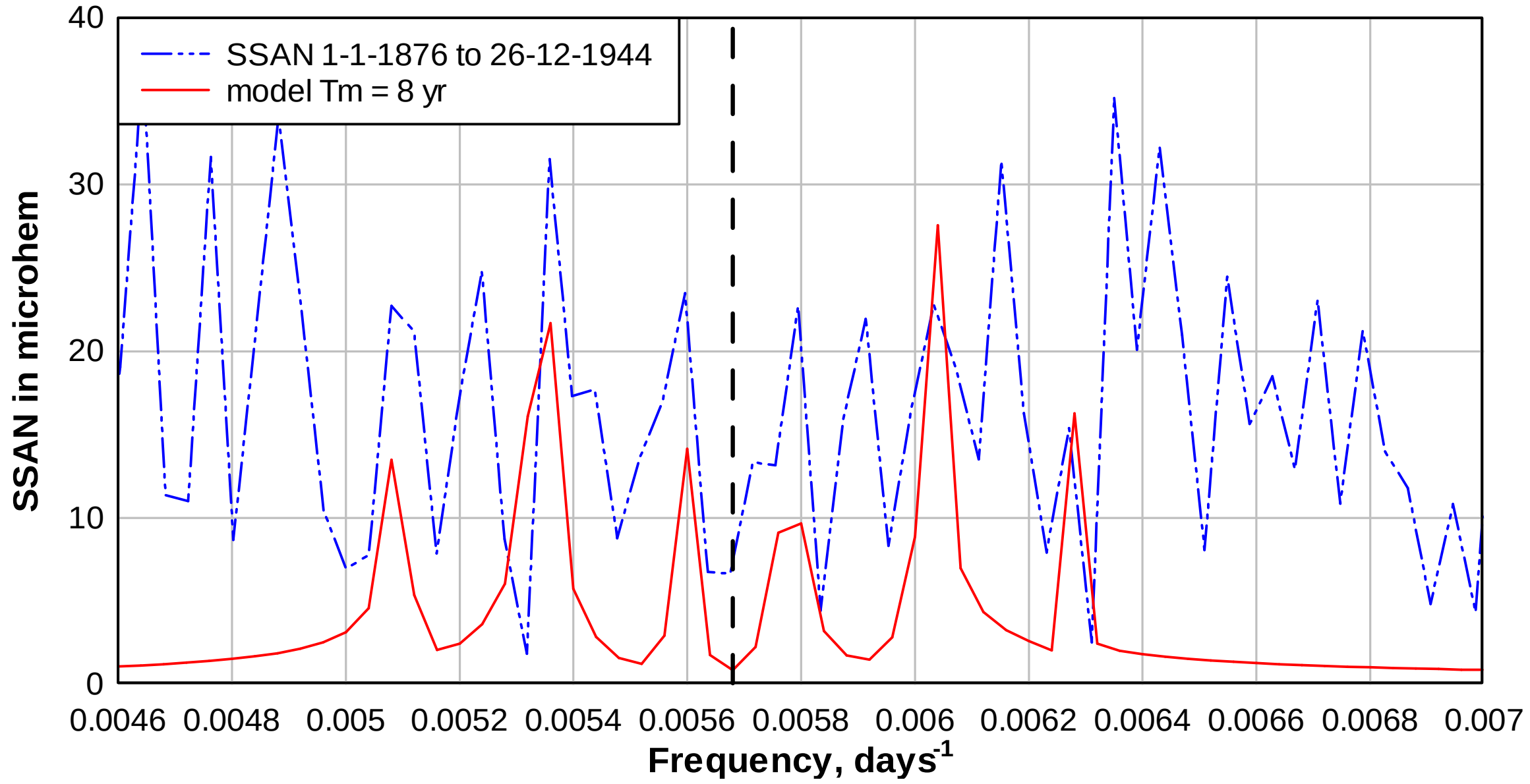

**Figure 17.** The model spectrum associated with the sub harmonic at 176 days when A = 0, the episode modulation period $T_m$ is 8 years and the solar cycle modulation period is 11 years. Also, shown for comparison, the spectrum of the first half of the SSAN data spanning 1876 to 1944.

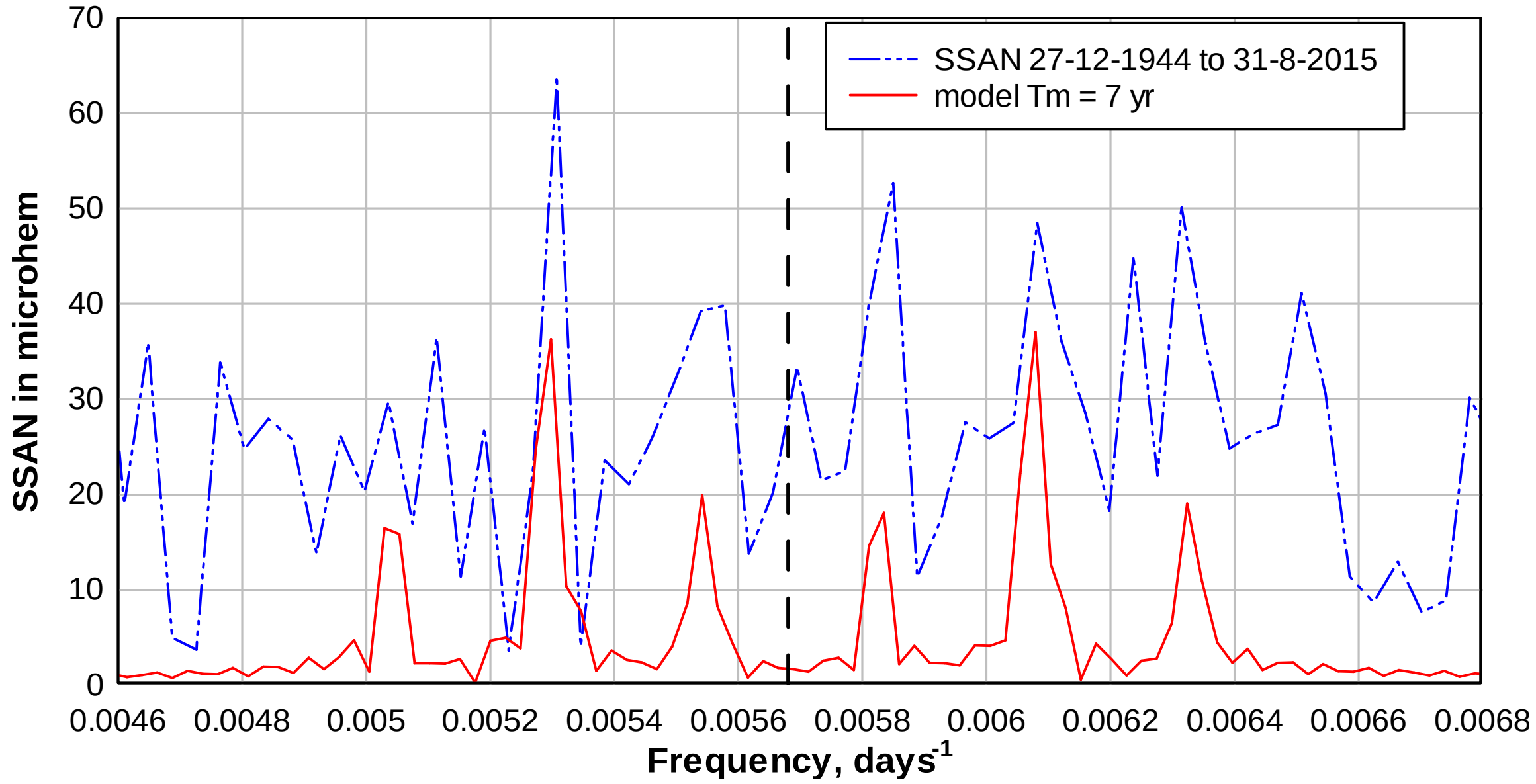

**Figure 18.** The model spectrum associated with the sub harmonic at 176 days when A = 0, the episode modulation period $T_m$ is 7 years and the solar cycle modulation period is 11 years. Also, shown for comparison, the spectrum of the second half of the SSAN data spanning 1945 to 2015.

By averaging the spectra from the first and second half of the record, Figure 19, the reduction in noise allows a model fit, again by eye, to be made with a higher level of certainty. The six prominent peaks in the average spectrum correspond reasonably closely with the spectral peaks obtained with a model with A = 0, $T_m$ = 7.5 years and $T_n$ = 11 years.

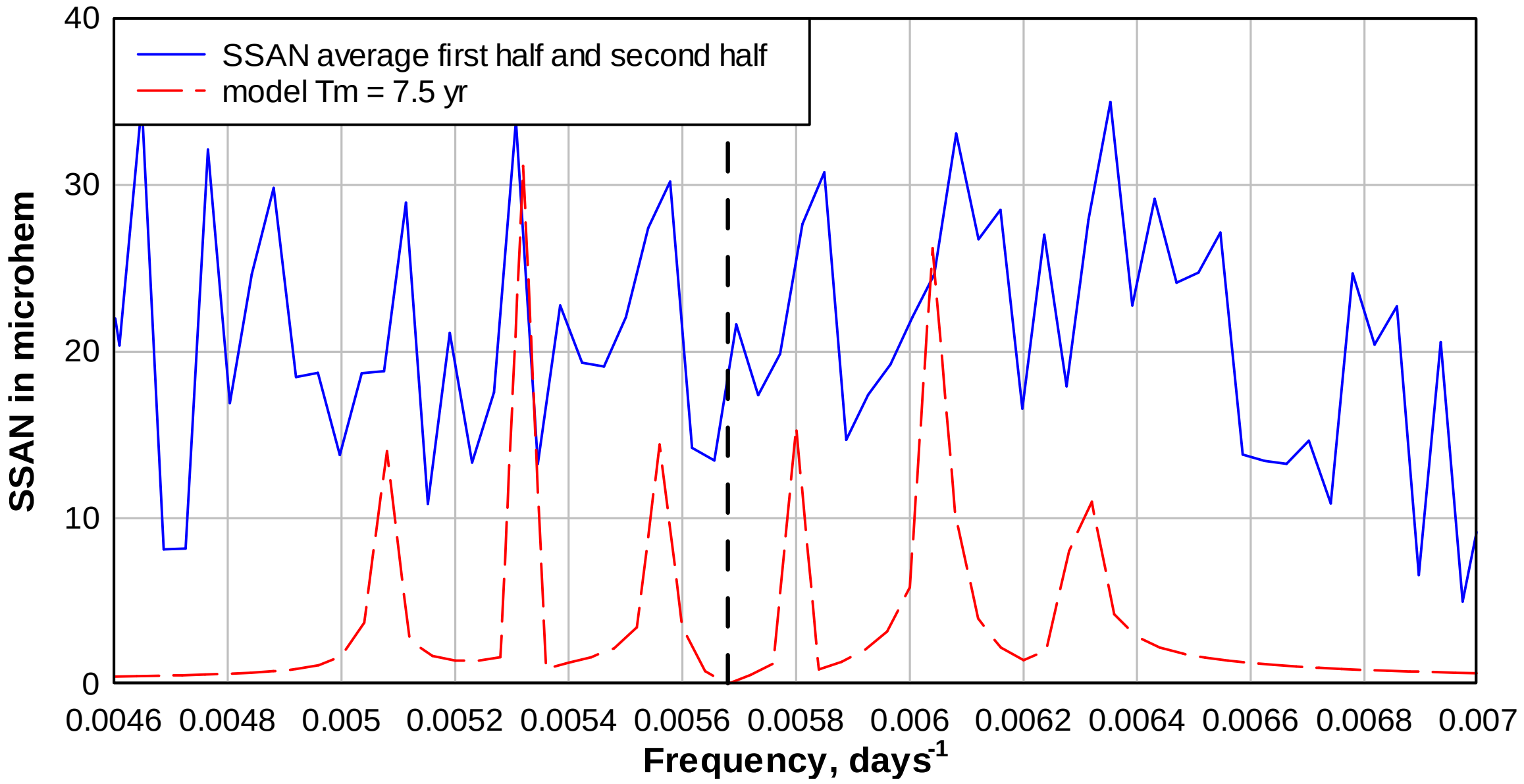

**Figure 19.** The model spectrum associated with the sub harmonic at 176 days when A = 0, the episode modulation period $T_m$ is 7.5 years and the solar cycle modulation period is 11 years. Also, shown for comparison, the average of the spectra of the first and second halves of the SSAN data.

As we are attempting a model fit to a spectrum that is expected to be symmetrical about the sub-harmonic frequency, $f_2 = 1/176 = 0.00568$ days$^{-1}$, a further reduction in noise can be obtained by folding the spectra about the point in the observed spectrum closest to this frequency and averaging, Figure 20.

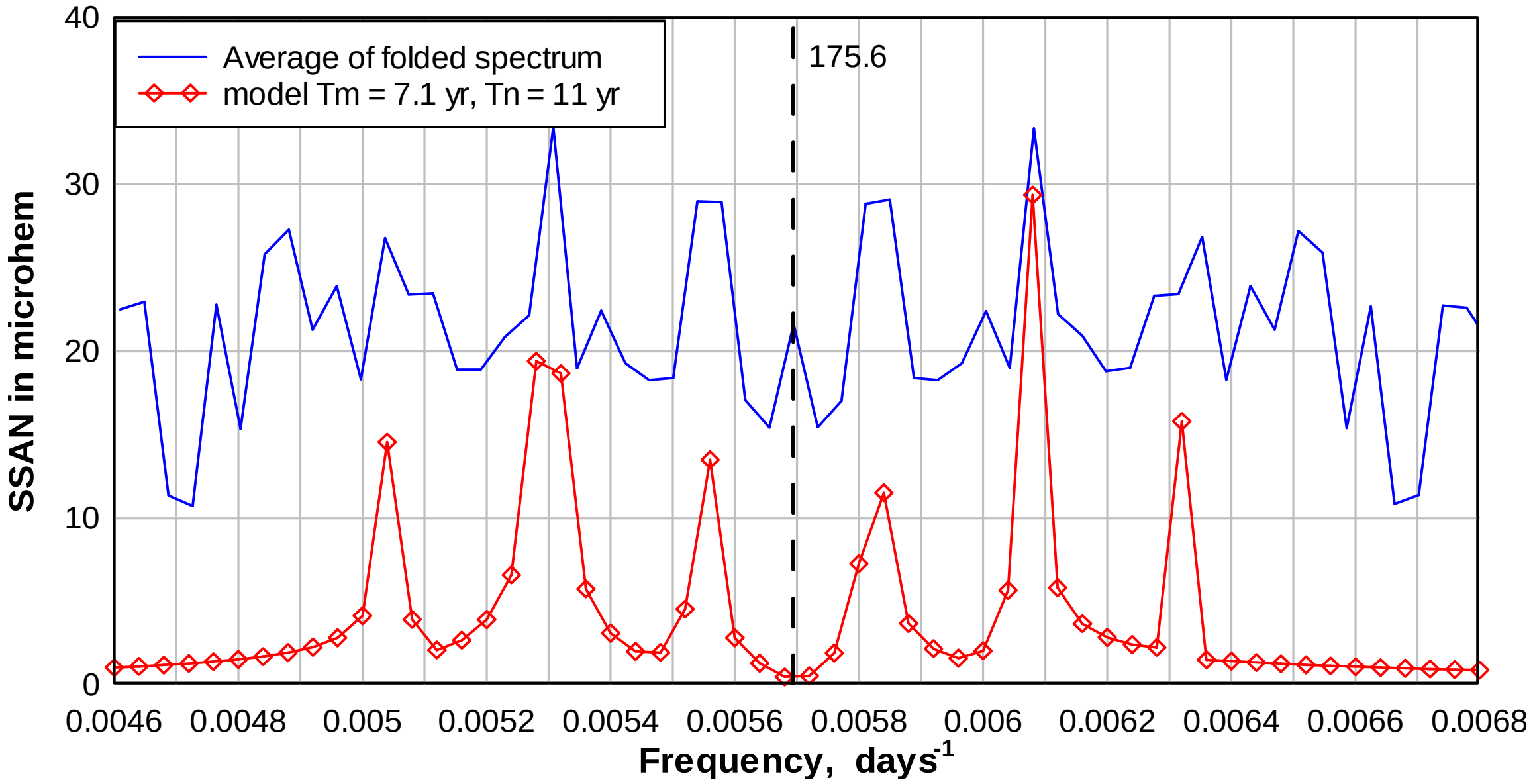

**Figure 20.** The model spectrum associated with the sub harmonic at 176 days when A = 0, the episode modulation period $T_m$ is 7.1 years and the solar cycle modulation period $T_n$ is 11 years. Also, shown for comparison, the folded average of the average of the spectra of the first and second halves of the SSAN data.

With $T_2$ = 176 days, $T_m$ = 7.1 years and $T_n$ = 11 years in the model the corresponding lower three sideband periods are at 158, 165, and 172 days. It is interesting to note that Lean and Brueckner (1989) found significant peaks at 155, 162 and 173 days in the spectra of sunspot area data spanning solar cycles 19, 20 and 21.

A similar approach can be used to fit a model spectrum to the spectrum of SSAN data in the spectral range around the 88 day periodicity of Mercury. Due to better resolution it is possible, in this spectral range, to average the spectra taken over four equal length subsets of the sunspot area data between 1876 and 2015. The average is shown in Figure 21. Model fits for $T_n$ = 11 years and $T_m$ = 3.5, 3.8 and 4 years are also shown.

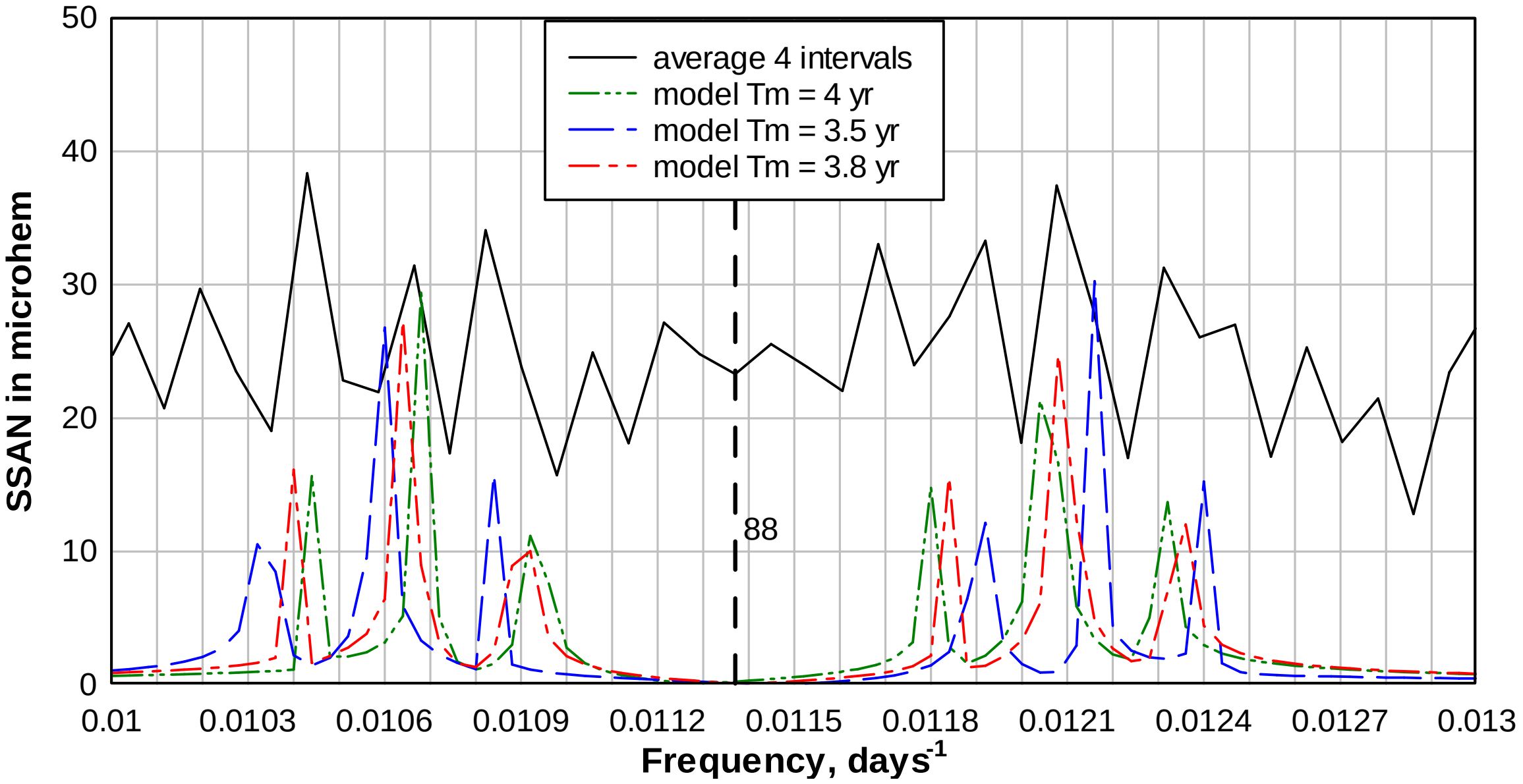

**Figure 21.** The average of the spectra from four equal length subsets of the SSAN data 1876 to 2015 compared with model spectra associated with the 88 day periodicity when A= 0, $T_m$ is 3.5, 3.8, and 4 years and $T_n$ is 11 years.

Using the folding and averaging technique on the data in Figure 21 we obtain Figure 22 and a fit to the observed spectrum of a model with $T_m$ = 3.8 years and $T_n$ = 12 years. The model sidebands are at 81.2, 82.7, 84.3, 92.0, 94.0 and 96 days.

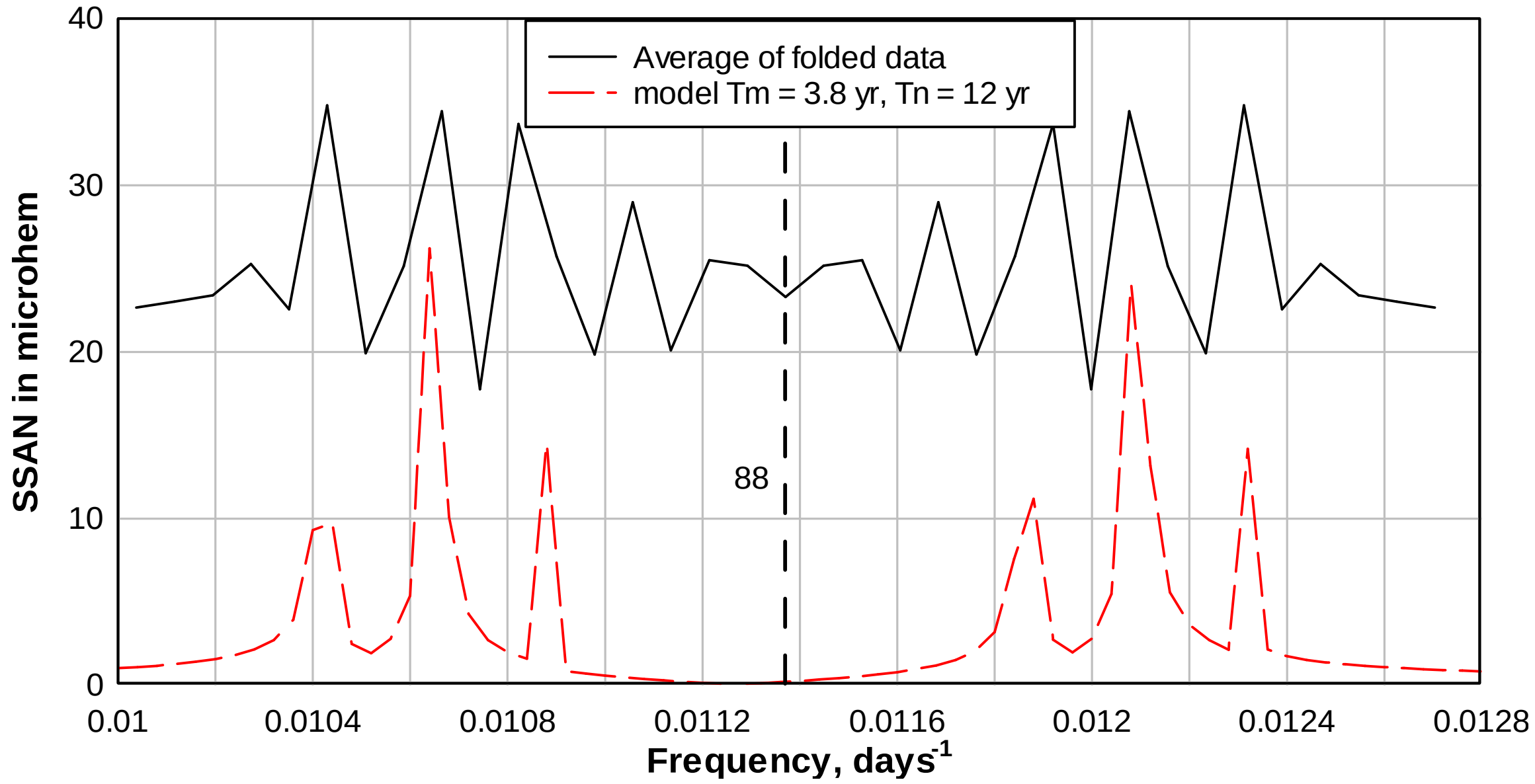

**Figure 22.** The folded average of the average spectrum from four equal length subsets of the SSAN data compared with a model spectrum associated with the 88 day periodicity when A = 0, $T_m$ = 3.8 years and $T_n$ = 12 years.

The model time variation with A = 0, $T_1$ = 88 days, $T_m$ = 3.8 years and $T_n$ = 12 years is shown in Figure 23. It is interesting to compare the model variation shown in Figure 23 with the observed variation of the ~88 day component of SSAN during solar cycle 23, Figure 3, and to note the resemblance between the modelled and observed variations.

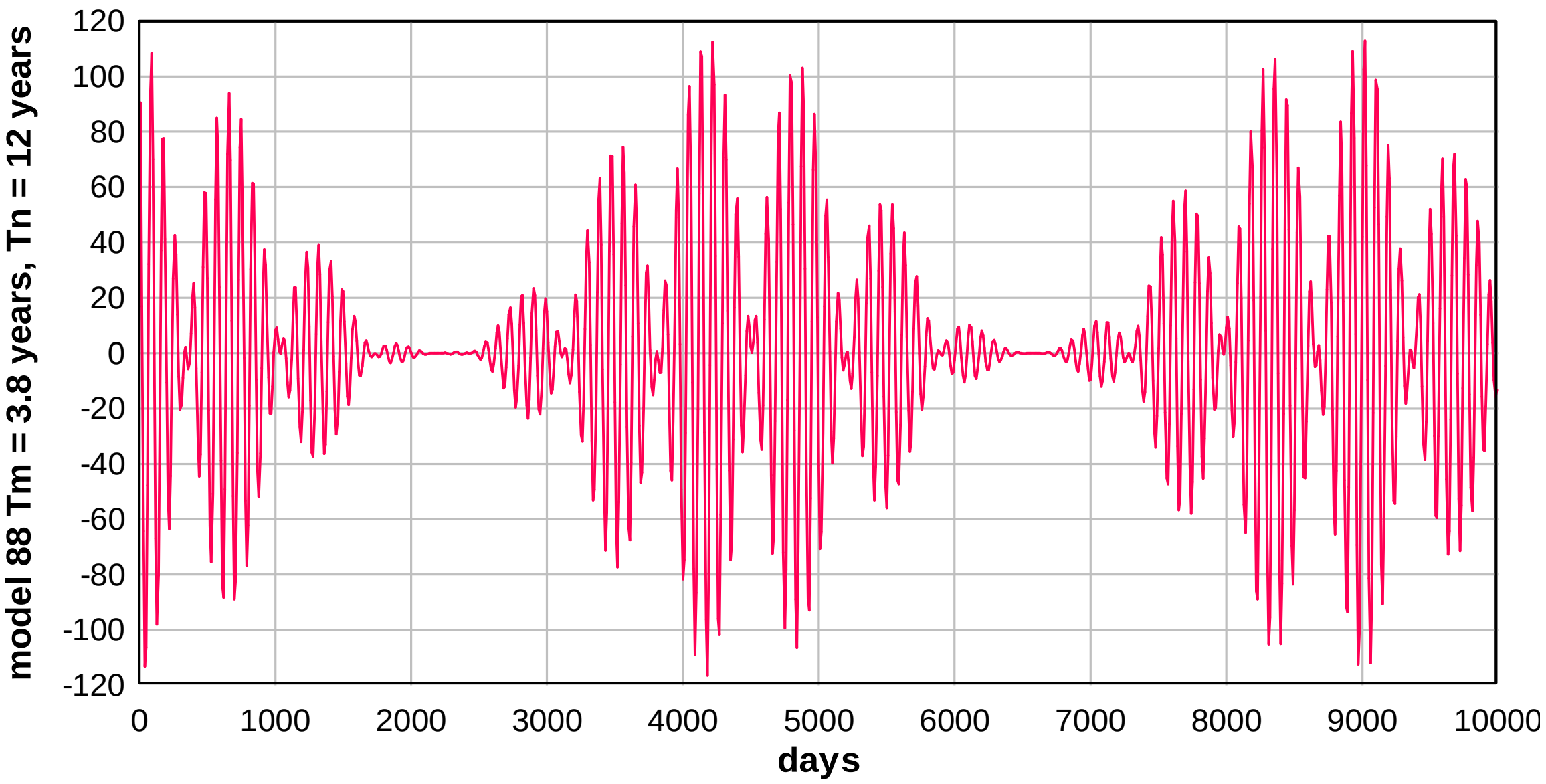

**Figure 23.** The time variation of the model with A= 0, $T_1$ = 88 days, $T_m$ = 3.8 years and $T_n$ = 12 years. The resemblance to the observed variation of the ~88 day component of SSAN during solar cycle 23, Figure 3, is close.

We conclude this section by calculating a model variation comprising the combined average of several sub harmonic model variations. The objective is to assess whether a model fit to the entire intermediate period spectrum of sunspot area can be obtained with a combination of the contributions from the higher sub harmonics. As indicated in Figure 12, with variations from sub harmonics higher than the first, 176 day, sub harmonic the modelled spectra are expected to overlap. As a result, a visual fit of observed spectrum to a simulated spectrum associated with a single sub harmonic, as achieved above for the 176 day spectrum, is not possible for a spectrum with contributions from higher sub harmonics. Therefore it is necessary to attempt a fit between the observed spectrum and the very complex, overlapping model spectrum associated with multiple sub harmonic contributions. There are three parameters of the model of equation 4 that can be varied to obtain a fit: A, which is an inverse measure of modulation strength and has been taken as zero previously, $T_n$, the solar cycle period, and $T_m$ the episode modulation period. To reduce the number of parameters to fit we set $T_n$ to 11 years. We set $T_m$ to 5.4 years which is the average of the 3.8 years and 7.1 years previously fitted, visually, to the 88 day and 176 day spectrums, respectively. $T_m$ set to 5.4 years corresponds to an average interval between sequential discrete episodes of 2.7 years. As we wish to compare the simulated spectrum with the observations of Lean (1990) which cover the period range between 50 and 500 days we limit the number of sub harmonics considered to 9 i.e. we consider contributions in the range from 88 days to 10 x 88 = 880 days. This leaves A, the modulation strength, as the single parameter to vary to obtain a fit. Note that the modelled spectrum will be an average over the total 90 peaks associated with 9 peaks in each of the ten modelled spectra. We first obtain an average frequency spectrum, smooth this with a five point running average then convert to a periodogram. Also note that we do not expect a good fit at that part of the spectrum associated with the 88 day period and 176 day sub harmonic period as we are using an average $T_m$ of 5.4 years rather than the best fit $T_m$ = 3.8 years for the 88 day spectrum and $T_m$ = 7.1 days for the 176 day spectrum. However, this section is interested in the spectrum range where the individual sub harmonic spectrums overlap. With $T_m$ = 5.4 years and $T_n$ = 11 years we find, fitting by eye, that a modulation strength A = 0.2 provides a reasonable fit to observations in this spectral range, Figure 25. Before discussing the observed and modelled spectrums we note that the average model time variation due to the ten combined contributions, Figure 24, evidences the ~11 year solar cycle but otherwise appears random. The spectrum of this variation, calculated over 25,000 days, provides the model periodogram of Figure 25.

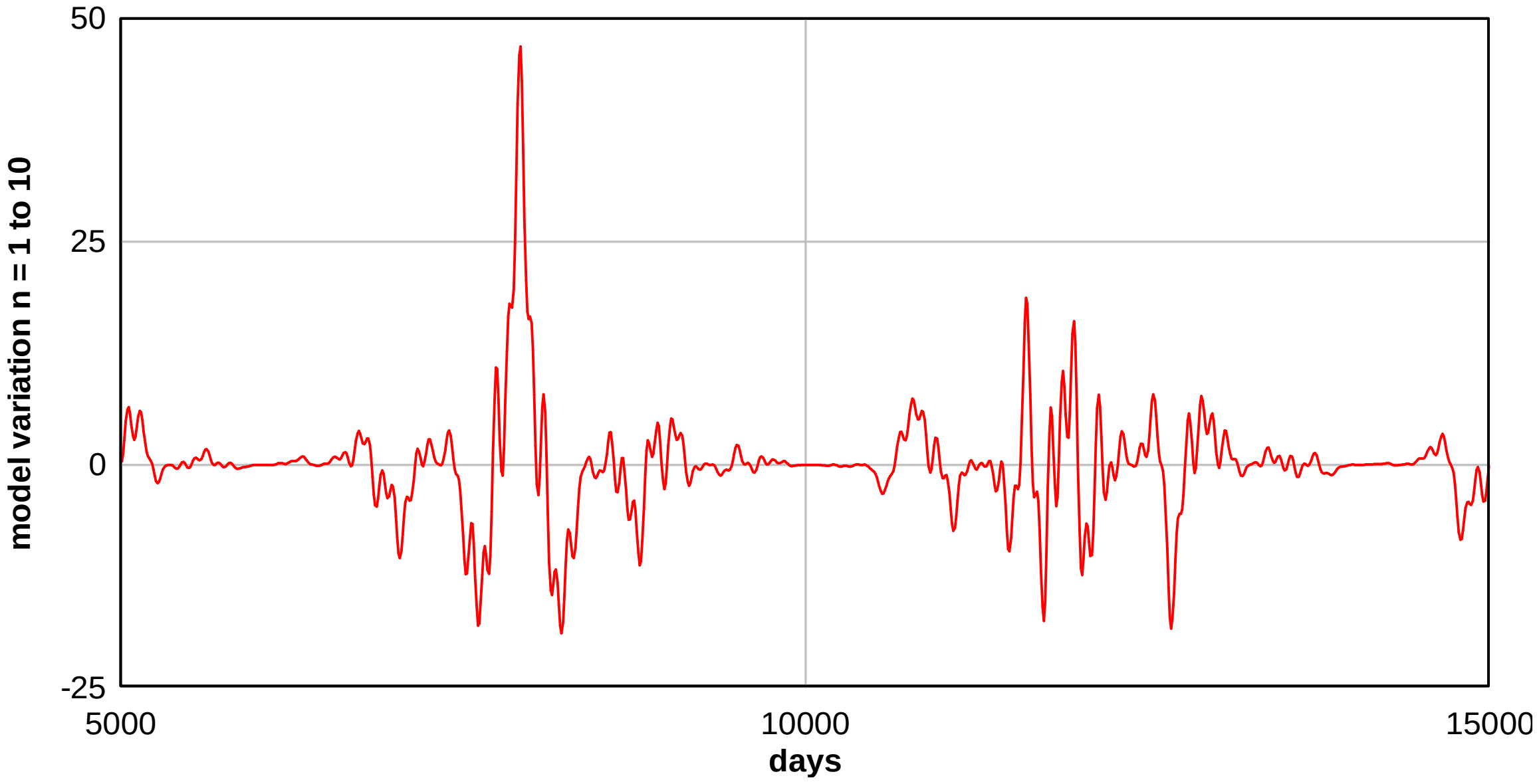

**Figure 24.** Illustrates the time variation associated with a model comprising the average of the 88 day period variation and nine sub harmonic period variations, period range 88 to 880 days.

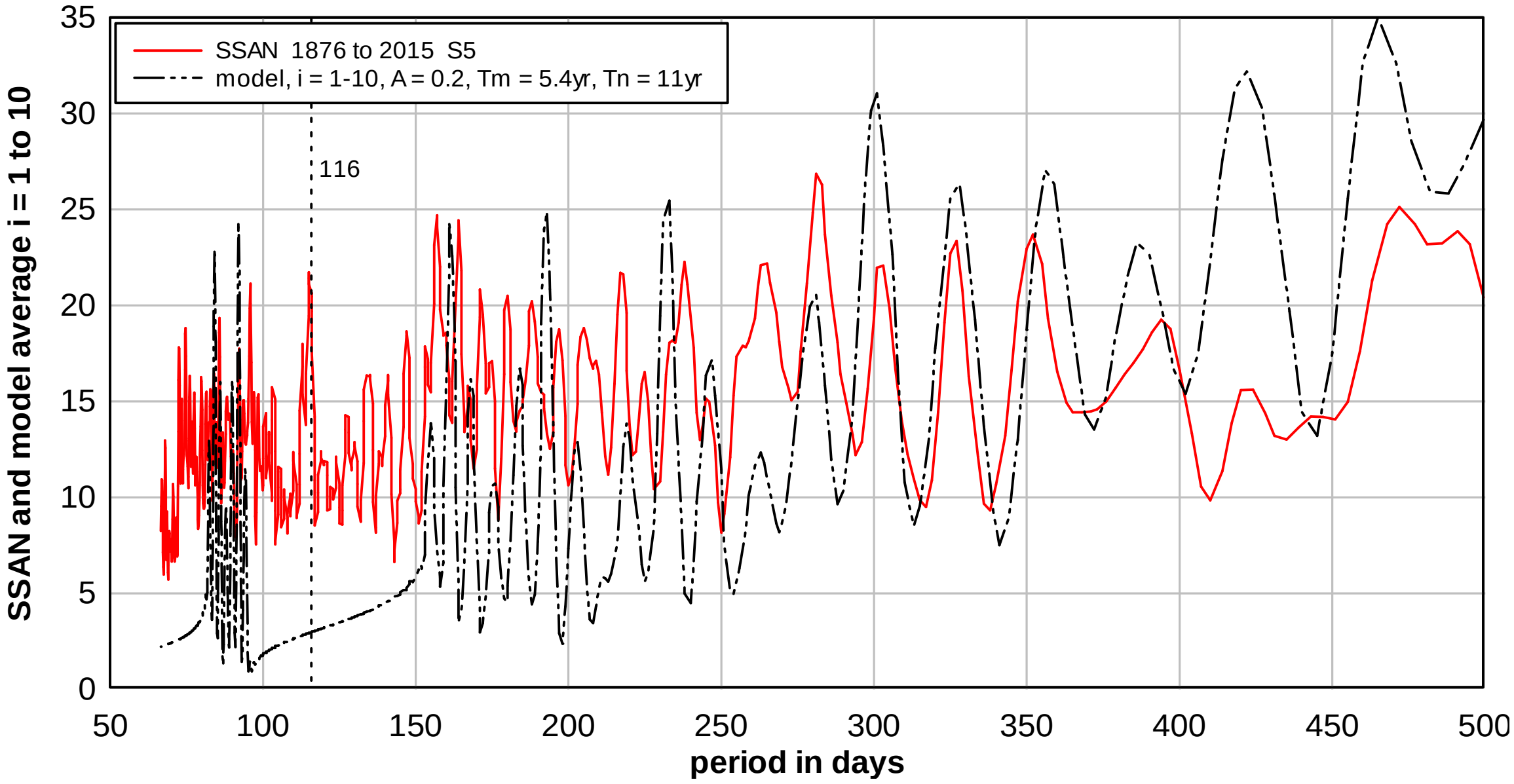

**Figure 25.** The model periodogram obtained from an average of the model variations associated with the Mercury periodicity and the sub harmonic periodicities in the range 88 to 880 days. The average spectrum has been subjected to 5 point smoothing. Also shown, full line, is the periodogram of the SSAN data 1876 to 2015 after 5 point smoothing.

Referring to Figure 25 we note that, in the range from 200 days to 500 days there is excellent agreement between periods of the observed peaks and the periods of the modelled peaks. That this modelled spectrum provides a close fit with the observed spectrum of SSAN in this substantial fraction of the intermediate periodicity range provides compelling support for the influence of Mercury on sunspot activity.

The general features of the modelled periodogram of Figure 25 can be compared directly with the observed periodogram of SSAN obtained by Lean (1990), which covers the same periodicity range and is presented as Figure 2C in that article. While the fine details are not expected to coincide as Leans SSAN data extends to 1987 whereas the SSAN data used here extends to 2015, the general features of the modelled periodogram in the period range 200 to 500 days and Leans observations in the same range are similar.

Lean (1990) averaged spectra calculated over three separate 31 year segments of the SSAN data available at that time. When we average the SSAN data over the three separate 47 year segments available now we obtain Figure 26.

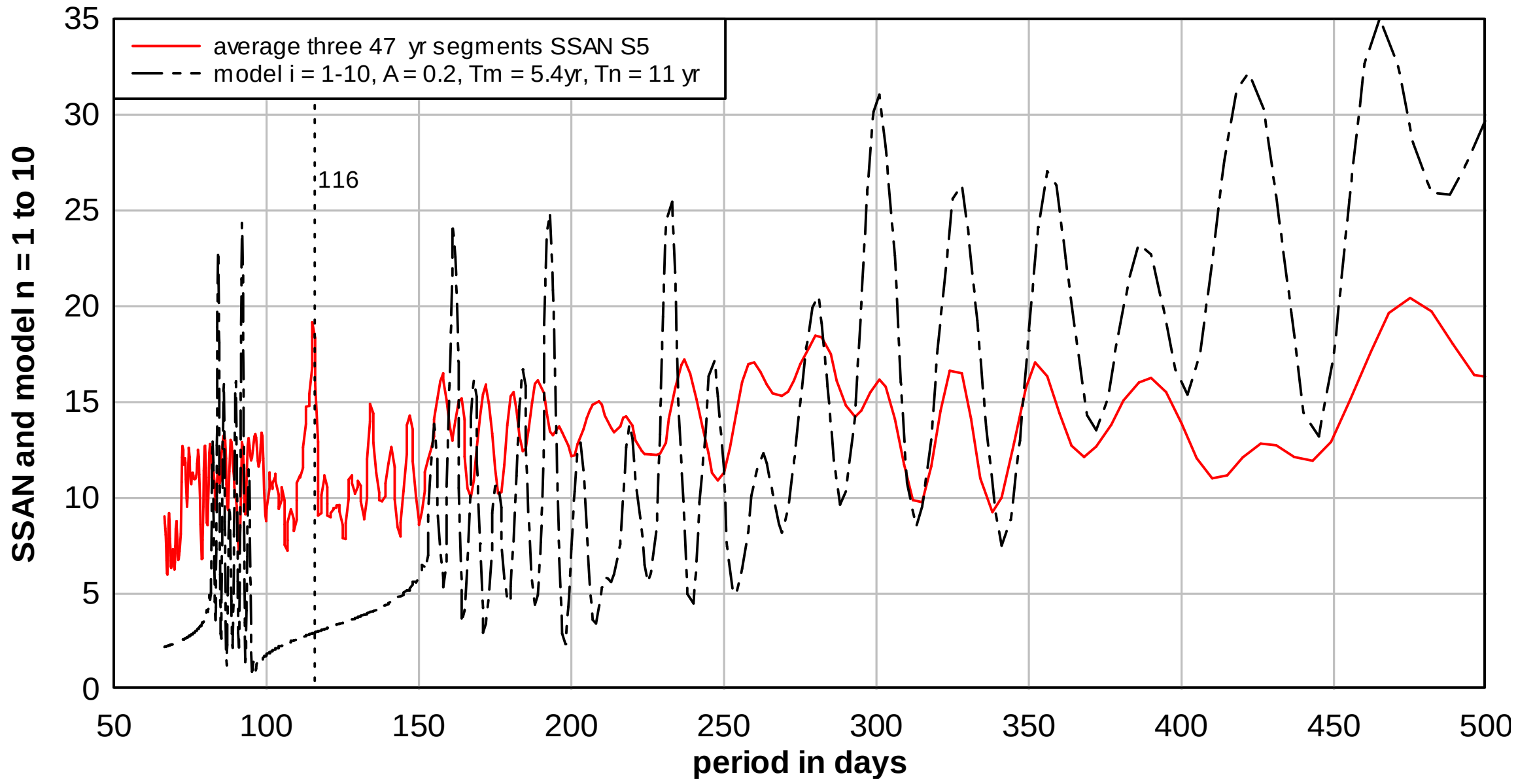

**Figure 26.** The model periodogram obtained from an average of the model variations due to Mercury periodicity and sub harmonic periodicities in the range 88 to 880 days after 5 point smoothing. Also shown is the periodogram obtained from the average of the frequency spectra obtained from three 47 year subsets of the SSAN data 1876 to 2015 after 5 point smoothing.

The intensity of the observed peaks in the three segment average is reduced, indicating considerable variation between spectra of each of the three segments, however, each of the eight peaks at the long period end of the observed spectrum still correspond closely, in period, with the periods of the eight peaks at the long period end of the modelled spectrum. This supports the selection of the parameters chosen to fit the model spectrum to the observed spectrum. In particular the episode modulation period, $T_m$ = 5.4 yr, although selected as an average from the fit to the individual 88 day and 176 day spectra, appears to provide a good fit to the longer period part of the SSAN spectrum. The position of the modelled peaks in the long period half of the periodogram is the result of the overlapping of individual sub harmonic sideband contributions and is, therefore, related very indirectly to sub harmonic periodicity. This indirect relationship is examined in detail below. To assess the stability of the observed spectrum we compare the spectra of SSAN for each of the three 47 year segments of SSAN data, Figure 27.

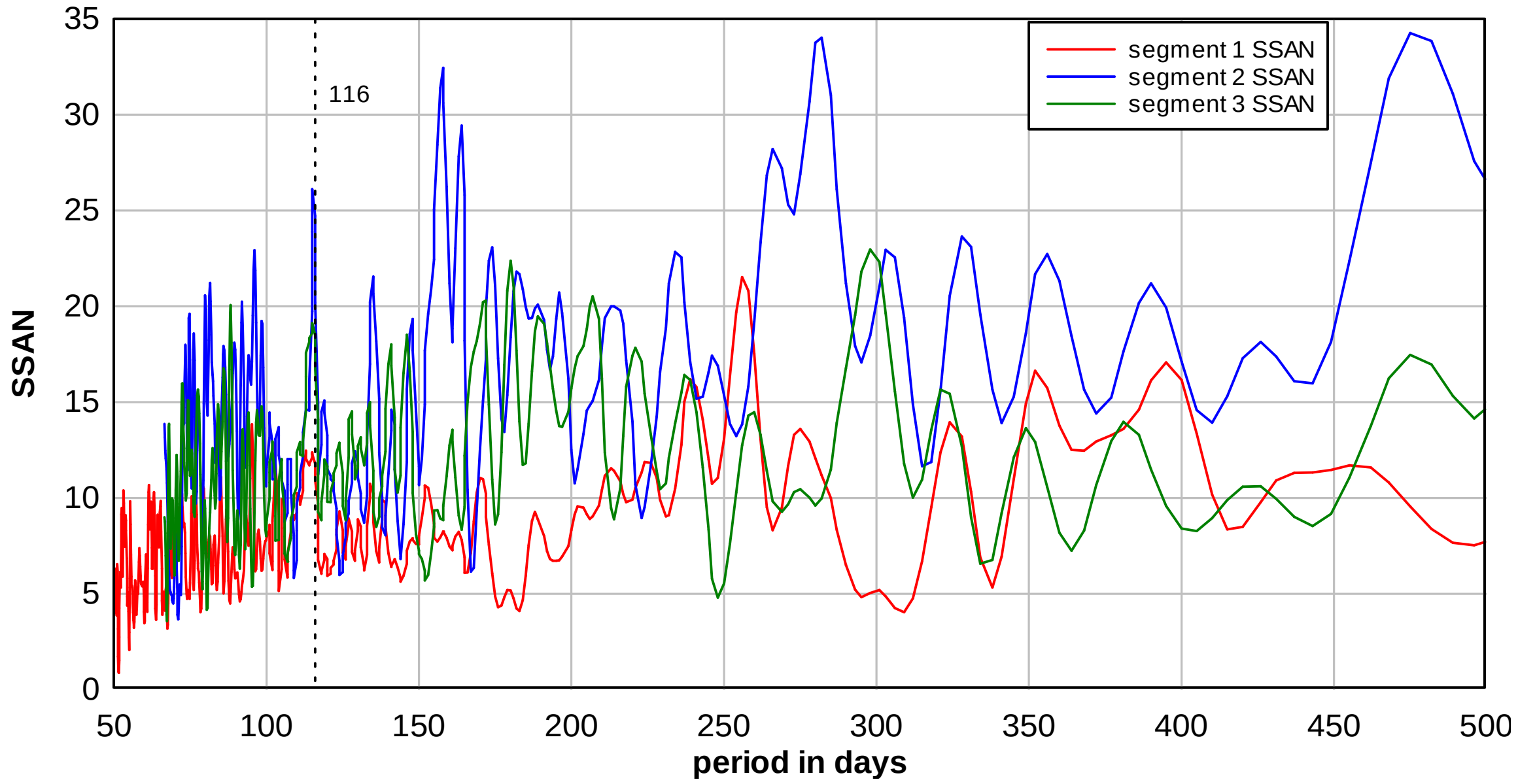

**Figure 27.** Periodograms obtained from each of the three 47 year segments of the SSAN data 1876 to 2015. The stability at the longer period end of the periodograms is evident.

It is clear that there is a large amount of variability between spectra of each segment. This is expected because, as shown above, the sidebands depend strongly on the episode modulation period, $T_m$, and, as shown in Figure 2, $T_m$ varies from solar cycle to cycle. However, it is interesting to note that the positions of the peaks in the longer period end of the spectrum are relatively stable. We now examine in detail the position of the peaks in the longer period part of the spectrum. To facilitate this we revert to consideration of frequency spectra and, in Figure 28, show the average of the modelled spectra for the first ten sub harmonics including the 88 day fundamental.

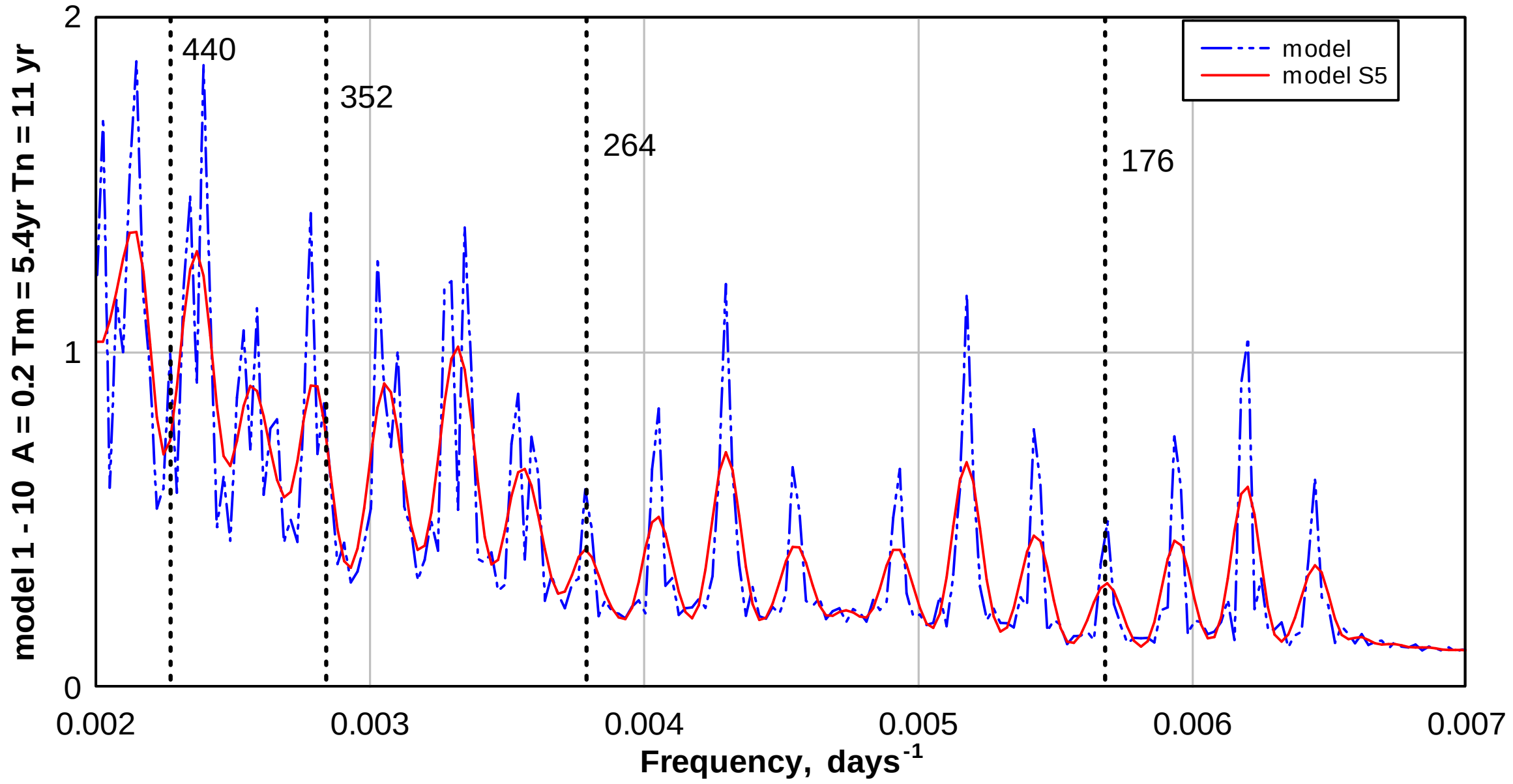

**Figure 28.** Frequency spectrum of the model variation comprised of the average of model variations associated with Mercury periodicities between 88 and 880 days. Reference lines at frequencies corresponding to a few of the sub harmonic periods are also shown.

The first two model spectra, centred on 88 days and 176 days do not overlap. Higher sub harmonic spectra overlap beginning with the spectrum for the 264 day sub harmonic. Due to the overlapping the average spectrum exhibits an approximately regular progression of peaks at the lower frequency end of the spectrum. Examination of the overlapping individual spectra (very complex with 90 overlapping peaks so not shown here) indicates that at regular frequency intervals several of the sidebands are closely superposed and it is this that leads to the regular variation in peak intensity in both the modelled and the observed spectrum, Figures 25, 26 and 28. Possibly the mathematics of this overlapping of spectra could be worked out. However, here we note that frequencies of the peaks in smoothed average spectrum can be found, quite accurately, from an empirical fit:

$$f_z = 1/264 - z/(11.5*365) \qquad (7)$$

where z = 0, 1, 2, 3 ….. denotes the number of the peak from the peak at 264 days. For example the sixth peak occurs at frequency 0.00236 days$^{-1}$, period 424 days. Clearly, in the lower frequency end of the spectrum, these peaks are not related directly to the sub harmonic periods, some of which are indicated in Figure 28 by reference lines.

The periods generated by equation 7 are 264, 282, 302, 325, 352, 385, 424, 472, 531, ….. days. Lean and Brueckner (1989) observed peaks in total sunspot area during solar cycles 19, 20 and 21 at 270, 287 and 323 days and assessed the probability of these peaks occurring due to noise of less than 1%. As these peaks are close to the 264, 282, 325 day peaks calculated by equation 7 it is possible that equation 7 is applicable to more global measures of sunspot activity than the hemisphere measure considered here, SSAN.

We now extend the averaging of the number of sub harmonic spectra to 14 including the fundamental at 88 days, i.e. we average model spectra associated with Mercury period and sub harmonic periods in the range 88 days to 1232 days. The averaged model spectrum, an average over 126 peaks, is scaled for comparison with the spectrum of the entire SSAN record 1876 to 2015 in Figure 29.

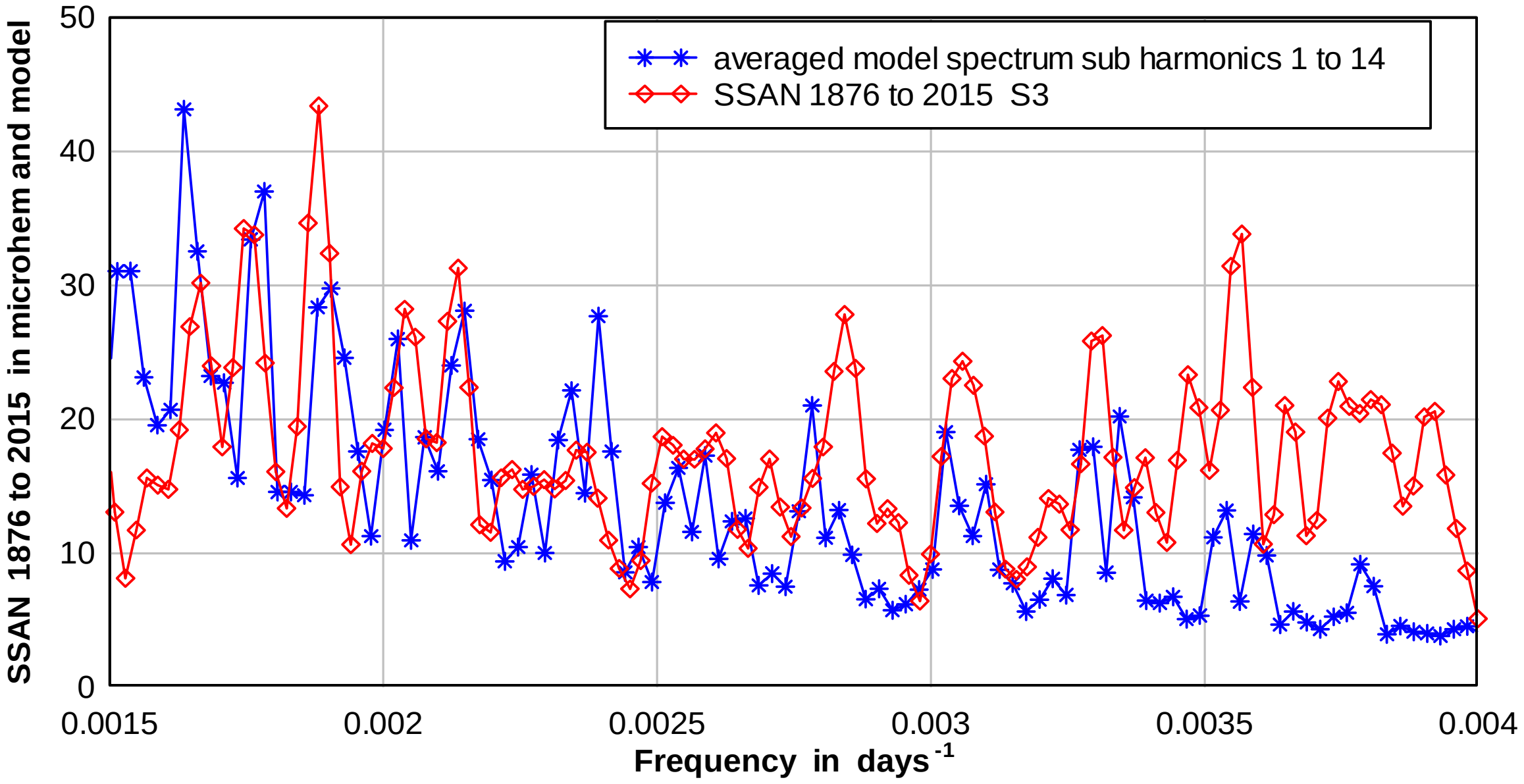

**Figure 29.** Spectrum obtained by averaging the fourteen individual model spectra associated with Mercury periodicities in the range from 88 days to 1232 days. The model parameters were A = 0.2, $T_m$ = 5.4 years, $T_n$ = 11 years. Also shown the spectrum of the SSAN data for the entire range 1876 to 2015 after 3 point smoothing. Note the excellent correlation of modeled and observed peaks in this substantial fraction of the intermediate spectral range.

There is a remarkably high correlation between the peaks in the observed spectrum and the twelve peaks in the modelled spectrum in the range 0.0015 days$^{-1}$ (666 days) to 0.004 days$^{-1}$ (250 days). Beyond these limits the correlation breaks down. However, the close correlation of nearly all the peaks over this substantial part of the intermediate periodicity range is compelling evidence that sunspot emergence is influenced by the variation of the radius of Mercury.

### 5.4 Sunspot emergence at the Mercury – Earth conjunction period.

In Figures 25, 26 and 27 we marked the spring tidal period of Mercury and Earth, $T_{ME}$ = 116 days, (more exactly, 115.89 days, Tan and Chen (2012)). Mercury and Earth are the most strongly elliptical inner planets. Figure 27 shows that the spectral peak in SSAN at 116 days is very stable. Unlike the Mercury sub harmonic periodicities, the $T_{ME}$ periodicity does not appear to have significant sidebands. From Figure 1 and equation 1 we note $T_{ME}$ is 1.005 of the allowed Rossby wave mode period T(2,9) = 115.32 days. Therefore Rossby wave response is likely. The occurrence and stability of the 116 day periodicity in SSAN data adds further evidence of the influence of Mercury on sunspot activity. However, further examination of 116 day periodicity is outside the scope of this article.

## 6. Discussion

A principal finding of this article is that the ~88 day component of sunspot area emergence occurs in episodes which, when discrete, are either exactly in-phase or exactly anti-phase to the 88 day time variation of the tidal effect of Mercury. Based on this observation and extending to consideration of sub harmonic periods, it was possible to derive an average model variation that had a spectrum closely matching the sunspot area spectrum in the intermediate range. Why phase inversion between discrete episodes of sunspot emergence occurs is unknown and here we speculate on a possible reason.

The hypothesis advanced in this article is based on prior evidence that the emergence of sunspots is related to the presence of magnetic Rossby waves on the Sun, (Lou 2000, Lou et al 2003, Dimitropolou et al 2008, Zaqarashvili et al 2010). We advanced the idea that Rossby waves, having mode periods very close to Mercury's 88 day period and sub harmonic periods, may have been stimulated to grow by the long term influence of the tidal effect of Mercury. We noted that the rotation period of the solar surface at the equator, 25.1 days, was such that closet approach of Mercury to the Sun occurs above the same point on the solar equator every seven solar rotations thereby favouring, relative to other planetary periodicities, stimulation of stationary, equatorially trapped, Rossby waves by Mercury. The tidal displacement due to Mercury is of the order 1 mm. Rossby waves with periods close to 88 and 176 day have been observed on the Sun, Sturrock and Bertello (2010) and the vertical displacement due to a Rossby wave is expected to be about 70 km, Lou (2000). Thus long term stimulation, if it occurred, could provide a 70 million times amplification of the tidal effect. As closest approach by Mercury corresponds to maximum tidal displacement we expect that peak displacement of a stimulated Rossby wave would occur at the point on the surface below closest approach by Mercury and, for a mode p = 1 Rossby wave, also at a point on the opposite side of the Sun. This supposed relation of Mercury orbit to Rossby wave displacement is illustrated very schematically in Figure 30 where the red full line represents an axis projection of the peak displacement point on the surface of the Sun, the blue broken line represents the time variation of the tidal effect of Mercury and the green line of amplitude 0.5 represents an in-phase Rossby wave response. If we assume that maximum sunspot emergence is triggered at the point of maximum positive Rossby wave displacement then sunspot emergence will be in phase with the p = 1 mode Rossby wave mode and maxima in the variation of the ~88 day component of sunspot area emergence will occur on day 0 (point A), day 88 (point B) and day 176 (point C) in Figure 30. In this case sunspot emergence will vary in-phase with the tidal effect of Mercury and will occur in two surface regions of the Sun $180^{o}$ apart. These two regions of the Sun will progressively be depleted of unstable sub surface magnetic flux according to the Babcock (1961) model.

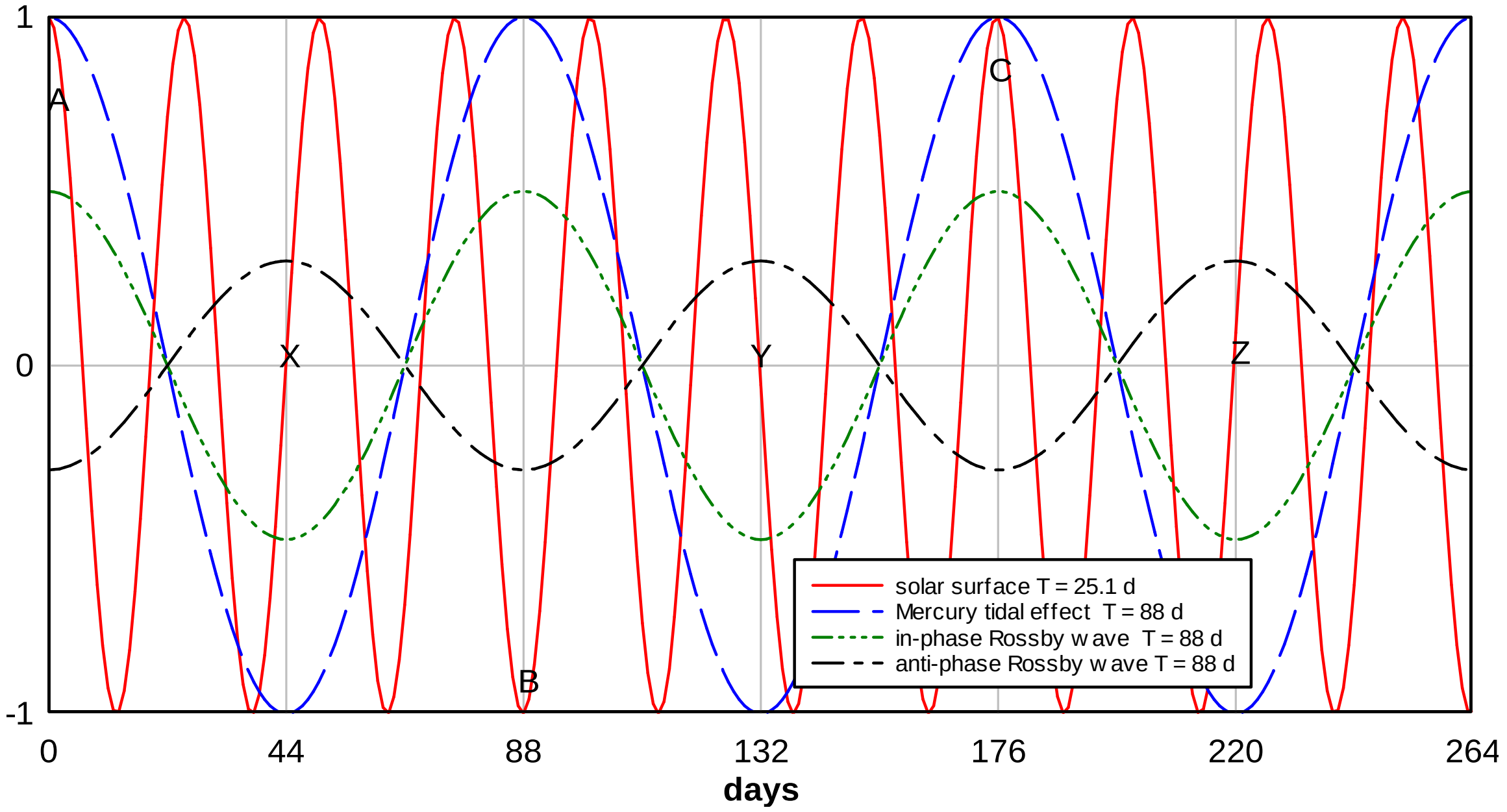

**Figure 30.** Illustrating a speculative scenario of the relationship between equatorial solar surface rotation at period 25.1 days, the Mercury tidal effect variation at period 88 days and long term Rossby wave mode variations at period 88 days on the Sun. The red full line represents the axial projection of a point on the solar equator directly below closest approach of Mercury. This point is associated with the maxima of the Mercury tidal effect, blue line. A p = 1, in-phase, Rossby wave mode will have peak displacement at a point directly below Mercury closest approach at times marked A and C and at time marked B at a point on the opposite side of the Sun. When unstable sub surface magnetic flux is depleted at these two points the depletion of magnetic flux by a p = 2 mode Rossby wave (black line) begins at two points shifted by $90^{o}$ on the solar surface from the original points. Depletion of sub surface magnetic flux and sunspot emergence now occurs at these points at times marked X, Y and Z. While on the diagram the time variations of the two Rossby wave modes are shown to occur simultaneously the sunspot emergence associated with each mode would occur in sequential episodes in-phase with either the p = 1 mode or the p = 2 Rossby wave and therefore exhibiting a $\pi$ phase change between one sequential episode and the next.

We speculate that when the two regions on the Sun are completely depleted of unstable magnetic flux a p = 2 mode Rossby wave, of the same period, begins to trigger sunspot emergence at points on the Sun with excess unstable subsurface magnetic flux. These are points displaced $90^{o}$ in longitude from the original two points. The ~ 88 day component of sunspot emergence will now have maxima on day 44 (point X), day 132 (point Y) and day 220 (point Z) and the phase of the ~88 day component of sunspot area will be shifted by $\pi$ relative to the phase of the Mercury tidal effect. Figure 30 indicates that the two Rossby wave variations occur at the same time. However, the variations in sunspot emergence would occur sequentially and would give rise to sequential discrete episodes of sunspot emergence. This scenario provides an explanation, albeit speculative, for the observed phase inversion of the variation of the ~ 88 day component of sunspot area between one discrete episode of sunspot emergence and the next. If this sequential process of unstable magnetic flux depletion and sunspot emergence continued in episodes lasting about two years it would generate several discrete episodes of the ~88 day component of sunspot area emergence similar to that observed in solar cycle 23 and reported in Figure 3.

This speculative scenario might explain other puzzling aspects of solar activity. It may explain why sunspots are not uniformly distributed in longitude but are centred at longitudes called active longitudes and why these active longitudes can persist for hundreds of years, Berdyugina and Usoskin (2003). It may explain the flip-flop phenomena where areas of sunspot emergence flip between one solar longitude and another at intervals of about two to four years, (Berdyugina and Usoskin 2003, Berdyugina 2006). Active longitudes in global rather than hemispheric measures of solar activity such as sunspot number or flare number can be seen, clearly, only for a few solar rotations e.g. Zhang et al (2015). A possible explanation is that active longitudes in global measures of solar activity will only be clearly evident when the phases of the ~88 day component or sub harmonic components of sunspot area are predominantly the same in both solar hemispheres. For example in section 3, comparison of Figure 3 with Figure 7 indicated that, during solar cycle 23, the ~88 day components of sunspot emergence were phase coherent in both hemispheres only in 1998 and 2004. Accordingly active longitudes should be clearly evident only in 2004, as observed by Zhang et al (2015), and in 1998.

The results of this paper suggest the possibility of predicting sunspot and flare activity. It is evident that once a discrete episode of sunspot emergence has begun in one hemisphere the strength of subsequent sunspot activity and therefore flare activity in that hemisphere is predictable to some extent. If, for example, it was established that a long, discrete episode of periodic sunspot emergence at the 176 day sub-harmonic period had begun in one solar hemisphere it is reasonable to expect flare activity in that hemisphere to vary with the same periodicity during the next few years. However, longer term prediction would be difficult despite the precision of the timing of the Mercury tidal effect. This is because the phase of the variation in succeeding episodes may be difficult to predict, the duration of an episode is quite variable and, as noted in Section 3, the phase and amplitude of the variations in the North and South hemispheres are often different. Further speculation is outside the scope of this paper.

## 6. Conclusions

For many years intermediate range quasi-periodicities in solar activity related variables have been studied by spectral analysis methods. However, previous studies did not attempt to establish phase coherence between the variation of solar activity related variables and a driving mechanism as, unlike the 11 year solar cycle and the 27 day solar rotation cycle, a driving cycle in the intermediate period range was unknown. In this paper the phase of the proposed intermediate period range driving mechanism, the tidal effect of Mercury, is known precisely. It has been possible to establish phase coherence between the tidal effect and the variation of the ~88 day and ~176 day components of sunspot area whenever successive episodes of the components were discrete. Phase coherence between two variables provides strong support of connection, more so than simply establishing spectral correlation. However, we have, in addition, been able to demonstrate excellent correlation between the observed spectrum of sunspot area data in the intermediate period range and model spectra based on episode and solar cycle modulation of sinusoids at the period of Mercury and its sub harmonic periods.

The observations presented above therefore support a connection between the tidal effect of Mercury and sunspot activity on the Sun. Based on these observations, past observations, and previous work concerning magnetic Rossby waves we can form a partial and speculative picture of a mechanism:

(1) There is a minute tidal elongation of the equatorial surface of the Sun due to the 88 day variation of the orbital radius of Mercury.

(2) Equatorial Rossby wave modes on the Sun that have allowed periods very close to the 88 day period of Mercury and to the sub harmonic periods of Mercury are excited by the tidal variation and, over a long interval, the Rossby waves grow to amplitudes that are sufficiently large to produce observable periodic variations in the Sun diameter.

(3) The Rossby waves trigger periodic emergence of loops of buoyantly unstable sub surface magnetic flux to form sunspots in episodes of periodic emergence that last for several years.

(4) An episode of periodic sunspot emergence lasts until the buoyant magnetic flux in the region of the Sun influenced by a particular Rossby wave mode is depleted.

(5) There may be between two and six episodes of sunspot area emergence associated with each Mercury periodicity during a solar cycle.

(6) When an episode is discrete in time from other episodes the variation of the sub harmonic components of sunspot area during the discrete episode are either exactly in-phase or exactly in anti-phase with the sub harmonic variation associated with the tidal effect of Mercury.

(7) Two states of the phase coherence have been made evident. The ~88 day component of sunspot area may be in-phase or in anti-phase with the tidal effect of Mercury. The

~176 day component of sunspot area may be in phase with every second tidal effect peak or in anti-phase with every second tidal effect minimum.

(8) When successive discrete episodes occur in the sub-harmonic components of sunspot area the spectrum of the sunspot area data exhibits sidebands with periods that depend strongly on the duration of the episodes. The periods of the sidebands in the spectrum associated with the Mercury sub harmonic of frequency $f_i$ can be obtained by a model of the form, model = $\cos(2\pi t f_i t)[A + \cos(2\pi t f_m t)](1 + \cos(2\pi f_n t))$, where $f_m$ and $f_n$ are the episode modulation and solar cycle modulation frequencies, respectively.

(9) A superposition of i = 1 – 14 model spectra with A = 0.2, $f_i$ = i/88 days$^{-1}$, $f_m$ = 1/(5.4x365) days$^{-1}$ and $f_n$ = 1/(11x365) days$^{-1}$ reproduces accurately the peaks in the low frequency range of the sunspot area spectrum indicating that a 2.7 year interval between episodes is typical in this periodicity range.

Some aspects arising from the present work are very speculative: (a), how the minute periodic variations in surface height on the Sun due to the tidal effect of Mercury can excite the long term growth of Rossby waves to observable amplitude; (b), how Rossby waves trigger the emergence of magnetic flux on the Sun; (c), why emergence occurs over varying intervals of a few years; (d), why, when intervals of emergence or episodes are discrete, there is a $\pi$ phase change in the periodic variation of sunspots from one episode to the next; and, (e), given that Mercury influences sunspot emergence, how do the other planets and conjunctions of planets influence sunspot emergence?

**References.**